\documentclass[traditabstract]{aa}

\usepackage{natbib}
\usepackage{graphicx,fleqn}
\usepackage{lscape}
\usepackage{times}

\newcommand\swift{{\em Swift}}
\newcommand\spitzer{{\em Spitzer}}
\newcommand\herschel{{\em Herschel}}

\def\lesssim{\mathrel{\hbox{\rlap{\hbox{\lower4pt\hbox{$\sim$}}}\hbox{$<$}}}}
\def\gtrsim{\mathrel{\hbox{\rlap{\hbox{\lower4pt\hbox{$\sim$}}}\hbox{$>$}}}} 

\begin{document}

\title{\herschel\thanks{\herschel\ is an ESA space observatory with science instruments provided by European-led Principal Investigator consortia and with important participation from NASA} observations of gamma-ray burst host galaxies: implications for the topology of the dusty interstellar medium}

\author{P.~Schady\inst{1}, S.~Savaglio\inst{1,2,3}, T.~M{\" u}ller\inst{1}, T.~Kr{\" u}hler\inst{4}, T.~Dwelly\inst{1}, E.~Palazzi\inst{5}, L.K.~Hunt\inst{6}, J.~Greiner\inst{1}, H.~Linz\inst{7}, M.J.~Micha{\l}owski\inst{8,9}, D.~Pierini\inst{10,11,12}\thanks{Visiting astronomer}, S.~Piranomonte\inst{12}, S.D.~Vergani\inst{13,14} \and W.K.~Gear\inst{15}}
\institute{$^1$ Max-Planck-Institut f{\"u}r Extraterrestrische Physik, Giessenbachstra\ss e, 85748, Garching, Germany\\
\email{pschady@mpe.mpg.de}\\
$^2$ Physics Department University of Calabria, 87036 Arcavacata di Rende, Italy\\
$^3$ European Southern Observatory, 85748 Garching, Germany\\
$^4$ ESO, Chile\\
$^5$ INAF - IASF Bologna, Via Gobetti 101, I-40129 Bologna, Italy\\
$^6$ INAF - Osservatorio Astrofisico di Arcetri, Largo E. Fermi, 5, 50125, Firenze, Italy\\
$^7$ Max-Planck-Institut f{\"u}r Astronomie, K{\"o}nigstuhl 17, 69117 Heidelberg, Germany\\
$^8$ Sterrenkundig Observatorium, Universiteit Gent, Krijgslaan 218-S9, 9000, Gent, Belgium\\
$^9$ SUPA\thanks{Scottish Universities Physics Alliance}, Institute for Astronomy, University of Edinburgh, Royal Observatory, Edinburgh, EH9 3HJ, UK\\
$^{10}$ Institut de Recherche en Astrophysique et PlanŽtologie, CNRS-UniversitŽ
Paul Sabatier Toulouse, 9 avenue du Colonel Roche, F-31028 Toulouse, France\\
$^{11}$ Centre de Physique des Particules de Marseille, Aix Marseille UniversitŽ, 168 avenue de Luminy, F-13009 Marseille, France\\
$^{12}$ INAF - Osservatorio Astronomico di Roma, Via di Frascati, 33, 00040 Monteporzio Catone, Italy\\
$^{13}$ GEPI - Observatoire de Paris, CNRS UMR 8111, Univ. Paris-Diderot, 5 Place Jules Jannsen, F-92190 Meudon, France\\
$^{14}$ INAF/Osservatorio Astronomico di Brera, via Emilio Bianchi 46, 23807 Merate (LC), Italy\\
$^{15}$ School of Physics and Astronomy, Cardiff University, 5 The Parade, Cardiff, Wales CF24 3YB, UK}

\date{Accepted: Received: }

\abstract
{Long-duration gamma-ray bursts (GRBs) are indisputably related to star formation, and their vast luminosity in gamma rays pin-points regions of star formation independent of galaxy mass, out to the epoch of re-ionisation. As such, GRBs provide a unique tool for studying star forming galaxies through cosmic time independent of luminosity. Most of our understanding of the properties of GRB hosts (GRBHs) comes from optical and near-infrared (NIR) follow-up observations, and we therefore have relatively little knowledge of the fraction of dust-enshrouded star formation that resides within GRBHs (if any), as traced through dust emission observed at far-IR wavelengths. Currently $\sim 20$\% of GRBs show evidence of significant amounts of dust along the line of sight to the afterglow through the host galaxy, and these GRBs tend to reside within redder and more massive galaxies than GRBs with optically bright afterglows. In this paper we present \herschel\ observations of five GRBHs with evidence of being dust-rich, targeted to understand the dust attenuation properties within GRBs better. Despite the sensitivity of our \herschel\ observations, only one galaxy in our sample was detected (GRBH~070306), for which we measure a total star formation rate (SFR) of $\sim 100~M_\odot$~yr$^{-1}$, and which had a relatively high stellar mass ($\log[M_\ast]=10.34^{+0.09}_{-0.04}$). Nevertheless, when considering a larger sample of GRBHs observed with \herschel, it is clear that stellar mass is not the only factor contributing to a \herschel\ detection, and significant dust extinction along the GRB sightline ($A_{V,GRB}>1.5$~mag) appears to be a considerably better tracer of GRBHs with high dust mass. This suggests that the extinguishing dust along the GRB line of sight lies predominantly within the host galaxy ISM, and thus those GRBs with $A_{V,GRB}>1$~mag but with no host galaxy \herschel\ detections are likely to have been predominantly extinguished by dust within an intervening dense cloud.
}

\titlerunning{GRB Host \herschel\ Observations}
\authorrunning{Schady et al.}
\maketitle
\begin{table*}
\begin{center}
\caption{GRB afterglow properties\label{tab:GRBAGprops}}
\begin{tabular}{@{}lccccccc}
\hline
GRB & Redshift & RA & Dec & Offset$^\dag$ & $A_{V,GRB}$ & $\beta_{OX}^\S$ & $N_{H,X}$ \\
 & & \multicolumn{2}{c}{(J2000)$^\ddag$} & (arcsec) & (mag) & & $10^{22}$cm$^{-2}$ \\
\hline\hline
000210 & 0.846$^{(1)}$ & 01:59:15.58 & $-$40:39:33.02$^{(6)}$ & $<0\farcs{2}^{(11)}$ & $-$ & $<0.54$ & $0.17\pm 0.04^{(17)}$ \\
000418 & 1.118$^{(2)}$ & 12:25:19.30 & $+$20:06:11.6$^{(7)}$ & $<0\farcs{4}^{(12)}$ & 0.4--0.9$^{(14,15)}$ & $-$ & $-$ \\
070306 & 1.496$^{(3)}$ & 09:52:23.31 & $+$10:28:55.26$^{(8)}$ & $<0\farcs{2}^{(13)}$ & ${5.5^{+1.2}_{-1.0}}^{(16)}$ & $<-0.08$ & ${2.5^{+0.3}_{-0.2}}^{(16)}$ \\
081109 & 0.979$^{(4)}$ & 22:03:09.72 & $-$54:42:39.5$^{(9)}$ & $<0\farcs{2}^{(13)}$ & ${3.4^{+0.4}_{-0.3}}^{(16)}$ & $<0.3$ & ${1.1^{+0.1}_{-0.1}}^{(16)}$ \\
090926B & 1.243$^{(5)}$ & 03:05:13.94 & $-$39:00:22.2$^{(10)}$ & $<0\farcs{6}^{(13)}$ & ${1.4^{+1.1}_{-0.6}}^{(16)}$ & $<0.4$ & ${2.2^{+0.5}_{-0.5}}^{(16)}$ \\
\hline
\end{tabular}
\end{center}
$^\ddag$ All positions are from the optical afterglow with the exception of GRB~000210, which corresponds to the X-ray afterglow position.\\
$^\dag$ Relative offset between the GRB afterglow position and the centre of the host galaxy.\\
$^\S$ GRB afterglow optical-to-X-ray flux spectral index as defined in \citet{jhf+04}. Those GRBs with afterglow spectral index $\beta_{OX}<0.5$ have lower optical-to-X-ray flux ratios than predicted by the standard afterglow synchrotron theory, and as such are referred to as `dark'.\\
{\bf References:} $^{(1)}$ \citet{pfg+02}; $^{(2)}$\citet{bbk+03}; $^{(3)}$\citet{jrw+08}; $^{(4)}$\citet{kgs+11}; $^{(5)}$\citet{fmj+09}; $^{(6)}$ \citet{pfg+02}; $^{(7)}$ \citet{ksm+00}; $^{(8)}$ \citet{rlt+07}; $^{(9)}$ \citet{dca+08}; $^{(10)}$ \citet{mfd09}; $^{(11)}$ This work; $^{(12)}$ \citet{bck+03}; $^{(13)}$ \citet{kgs+11}; $^{(14)}$ \citet{ksm+00}; $^{(15)}$ \citet{bdf+01}; $^{(16)}$ \citet{kgs+11}; $^{(17)}$ \citet{pfg+02}
\end{table*}

\section{Introduction}
\label{sec:intro}
Ever since the connection between long-duration GRBs and massive star formation was firmly established \citep{gvp+99,hsm+03}, much attention has been focused on using GRBs to trace the cosmic star formation history (SFH). However, such ambitions have been hindered by indications that GRBs may not be indiscriminate tracers of star formation \citep{kyb+09, bbp10,wp10,re12} \citep[but see][]{mkh+12} and that further factors may play a dominant role in regulating the GRB formation rate (e.g. metallicity, the initial mass function; IMF). The GRB ``collapsar" model predicts a metallicity threshold of $\sim 0.3$~Z$_\odot$, above which massive stars cannot collapse to form a GRB \citep{mw99,ln06}, and there are observations that similarly point to GRBs occurring within low-metallicity environments \citep{fls+06,wp07,mkk+08,lkb+10,gf13}. There are, however, examples of GRBHs with near-solar metallicity or above \citep[][]{gfk+09,lkg+10,srg+12,ekg+13}, and rather than metallicity, other environmental properties have been suggested to play a more critical role in the formation of GRBs, such as the SFR per unit stellar mass (i.e. specific SFR or sSFR) \citep[e.g.][]{msc+11,kw11,kfm+14,hpm+14} or a high interstellar medium density \citep{mhp+14}.

Most of our understanding of the environmental properties of GRBHs comes from photometric and spectroscopic observations of bright, dust-poor galaxies, which make up the bulk of the population at $z<1$ \citep{sgl09,gf13}, and for which data are more readily available. Over the last decade, great emphasis has been put on extending the GRBH detection rate out to higher redshifts, and to more dust-rich systems. The rapid, sub-arcsecond positions provided by the GRB-dedicated NASA \swift\ mission \citep{gcg+04}, along with the commissioning of numerous near-infrared (NIR) facilities for GRB follow-up observations, such as PAIRITEL \citep{bsb+06}, GROND \citep{gbc+08} and more recently RATIR \citep{bkf+12}, have resulted in an increase in the fraction of detected afterglows that are significantly dust-extinguished (i.e. rest frame $V$-band dust extinction $A_{V,GRB}>1.0$~mag) \citep[e.g.][]{gkk+11}. These reddened GRBs tend to reside within galaxies that are more massive and luminous than the more frequently observed hosts of GRBs with optically bright afterglows \citep{kgs+11,hpr+11,slt+12,plt+13}.

Although it is still debated whether the fraction of more massive, dustier hosts are as numerous as would be expected if GRBs directly follow the star formation activity \citep{plt+13, hpm+14}, their detection certainly does present challenges to the GRB {\em `collapsar'} model, which requires a progenitor star metallicity cut-off. Studying the abundance and properties of dust within GRB host galaxies is thus important not only in the use of GRBs as cosmological tools, but also for our understanding of the progenitors and environmental factors that produce these catastrophic explosions. There have now been several published works that looked into the properties of the more massive and dust-rich GRB host galaxies \citep[e.g.][]{kgs+11,hpr+11,plt+13,rks+12}. However, for the most part these investigations have not included observations of the host galaxy dust emission, which probe the obscured star formation and can contain a significant fraction of the galaxy energy density.

\begin{table*}
\caption{\herschel\ observations}\label{tab:obsLog}
\begin{center}
\begin{tabular}{@{}lcclcc}
\hline   
GRB Host & Instrument & Date/OD & OBSID & Duration (s) & On-source time (s) \\
\hline\hline
000210 & SPIRE & 2012-05-11/1093 & 1342245552 & 1135 & 296 \\
 & PACS & 2012-12-11/1308 & 1342256975, 76 & $2\times 840$ & $2\times 270$ \\
000418 & SPIRE & 2012-07-12/1156 & 1342247974 & 1135 & 296 \\
070306 & PACS & 2012-10-30/1266 & 1342254142, 43 & $2\times 2250$ & $2\times 720$ \\
081119 & SPIRE & 2012-05-10/1092 & 1342245514 & 1135 & 296 \\
 & PACS & 2012-10-16/1251 & 1342253509, 10 & $2\times 2250$ & $2\times 720$ \\
090926B & SPIRE & 2012-03-01/1022 & 1342239863 & 1135 & 296 \\
 & PACS & 2013-01-05/1333 & 1342258541, 42 & $2\times 2250$ & $2\times 720$ \\
\hline
\end{tabular}
\end{center}
\end{table*}

We have used the infrared \herschel\ satellite \citep{prp+10} to study the dust emission properties of a small sample of five GRBHs that we considered good candidates of containing appreciable amounts of dust. With a sample of five host galaxies, our aim is to reach a greater understanding of the dust properties within this dustier and important class of GRBHs, rather than to draw any quantifiable conclusions on the relation between GRBs and the cosmic SFR. Our \herschel\ PACS and SPIRE observations span the wavelength range from 100$\mu$m to 500$\mu$m, which for our sample (at $0.8<z<1.5$) and for typical dust temperatures of 35K provide good coverage of the peak of the thermal dust emission. \herschel\ observations thus enable the most accurate determination of the obscured star formation within GRBHs.

Our GRBH \herschel\ observations presented here, combined with the sample of GRBHs presented in \citet[][HPM14 from here on]{hpm+14}, as well as GRBH~980425 \citep{mhp+14} and GRBH~031203 \citep{sod+14}, provide a total sample of 23 GRBHs observed with \herschel. The GRBHs in HPM14 hosted GRBs with a range in visual extinctions, and we combine this sample with the GRBHs presented in this paper to investigate a larger GRB afterglow and host galaxy parameter space. In section~\ref{sec:datanlys} we summarise previous multi-wavelength observations taken of our GRBH sample, and we describe our \herschel\ observations and data reduction. In section~\ref{sec:rslt} we present the results from our spectral energy distribution (SED) data analysis, and we summarise the principle characteristic properties of our sample in section~\ref{sec:HostProps} and compare them with the literature. Finally, in section~\ref{sec:disc} we explore the implications of the \herschel\ GRBH detection rate on the origin of the GRB afterglow extinguishing dust. Our conclusions are summarised in section~\ref{sec:sum}. Throughout the paper, we assume a $\Omega_m=0.3$, $\Omega_\Lambda=0.7$ cosmology, with Hubble constant $H_0=70$km~s$^{-1}$~Mpc$^{-1}$.

\begin{figure}
\centering
\includegraphics[width=0.5\textwidth]{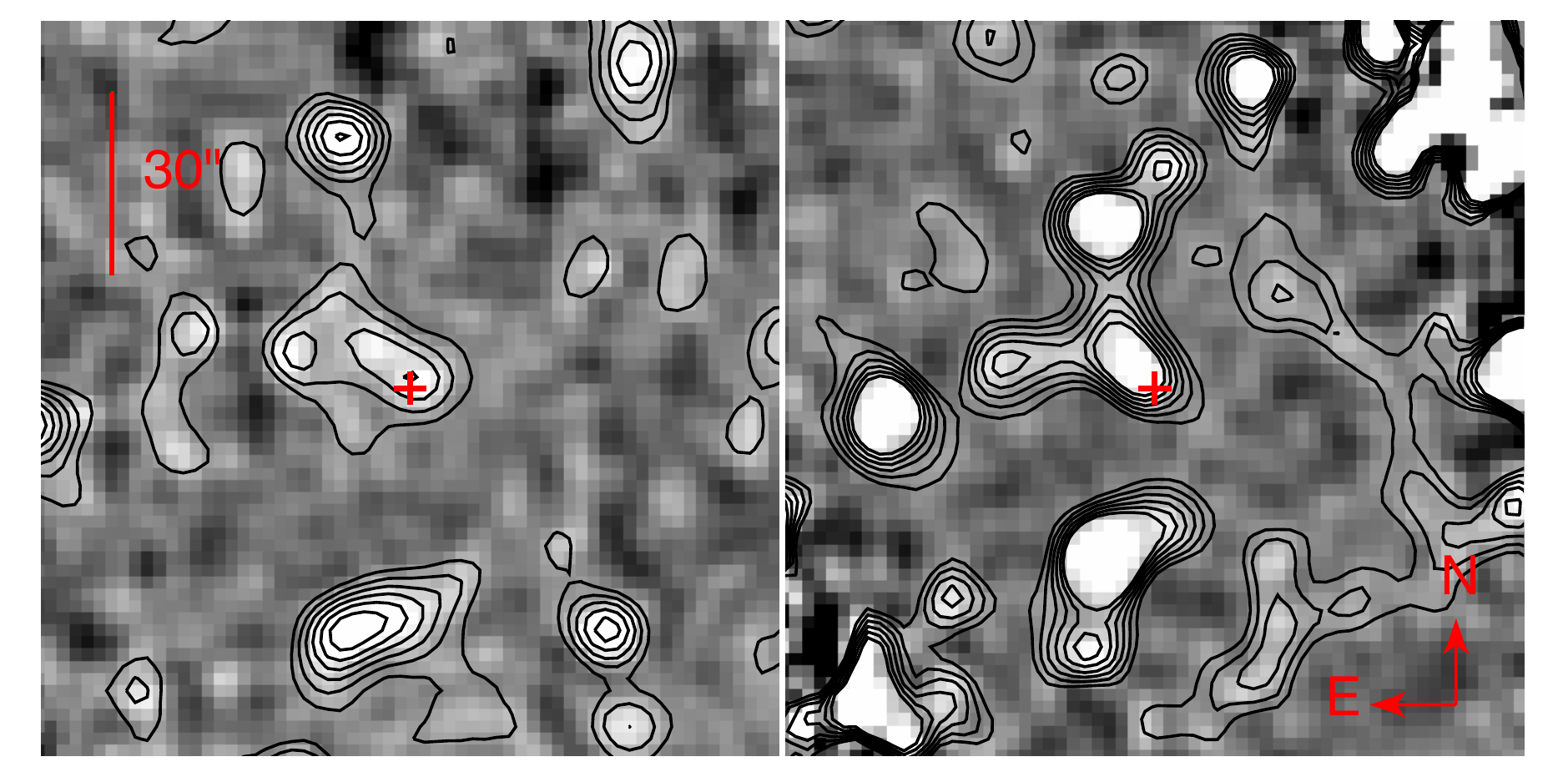}
\caption{\herschel/PACS 100$\mu$m and 160$\mu$m images of GRBH~070306. The images are 2\arcmin$\times $2\arcmin and 3\arcmin$\times $3\arcmin, respectively, centred on the GRB afterglow position (red cross), and have been smoothed using a Gaussian with $\sigma$ two times the pixel scale. The image is displayed with a linear greyscale ranging from $-1$~mJy~pix$^{-1}$ (black)  to $+1$~mJy~pix$^{-1}$ (white), and contour levels from 1$\sigma$ to 6$\sigma$ are over plotted in back. The spatial scale is indicated on the top left.}\label{fig:DETsample}
\end{figure}

\section{The sample}
\label{sec:datanlys}
Our host galaxy sample was built up using two selection criteria, both aimed at targeting those host galaxies with significant amounts of dust, and thus with a high chance of being detected with \herschel. Firstly, we selected the host galaxies of those GRBs with significantly dust-extinguished afterglows (i.e. $A_{V,GRB}>1$~mag) that were at $z<1.5$ and that had been observed at submillimetre (submm) wavelengths as part of our programme with the LABOCA instrument \citep{skk+09} on APEX \citep{gns+06}. By limiting our sample to redshifts $z<1.5$ we increase our chance of a detection due to the \herschel\ sensitivity limit, and the additional submm data provides constraints on the dust emission at wavelengths longward of the thermal emission peak. At the time of the submission of the \herschel\ proposal, this gave us a sample of three GRBHs. In addition to this sample of host galaxies, we also included those GRBHs for which dust emission had already been detected at 850$\mu$m with SCUBA on JCMT \citep{hrg+99}, which increased the sample by a further two host galaxies. All five host galaxies in our sample have been well observed at optical and NIR wavelengths, as well as with the {\em Spitzer Space Telescope}.

\subsection{Afterglow parameters}
The five GRB host galaxies in our sample are listed in Table~\ref{tab:GRBAGprops} together with some properties related to the GRB optical and X-ray afterglow. In addition to the afterglow visual extinction, $A_{V,GRB}$, we also list the afterglow optical-to-X-ray spectral index, $\beta_{OX}$ \citep{jhf+04}, which provides a diagnostic tool for identifying GRBs with optical afterglows dimmer than expected by GRB synchrotron emission theory. In those cases where the GRB redshift is known to be $z<4$ (i.e. the optical afterglow is not significantly absorbed by neutral hydrogen within the intergalactic medium), a spectral index $\beta_{OX}<0.5$ is an indicator of a dust-extinguished afterglow. The spectral index $\beta_{OX}$ is much easier to measure than $A_{V,GRB}$\footnotemark[1], although it doesn't quantify the amount of dust along the GRB sight-line. We discuss further the merits of $A_{V,GRB}$ and $\beta_{OX}$ at identifying GRB host galaxies appreciable levels of dust in section~\ref{sec:disc}.
\footnotetext[1]{Whereas an X-ray afterglow flux and $R$-band afterglow flux limit is sufficient to determine the $\beta_{OX}$ spectral index, an accurate determination of $A_{V,GRB}$ requires the GRB afterglow to be detected in the X-ray and at least four optical or NIR filters \citep{sdp+12}}

All \swift\ GRBs (GRB~070306, GRB~081109, GRB~090926B) had their afterglows clearly detected at optical and NIR wavelengths with ground-based facilities, and at X-ray wavelengths with the X-ray telescope onboard \swift\ \citep{bhn+05}. The GRB afterglow is typically well explained by synchrotron emission, and thus the extinction and absorption from dust and gas of the intrinsically simple afterglow spectrum (power law or broken power law) can be well measured from the afterglow SED, in particular when afterglow data at the mostly unattenuated NIR and X-ray ($\gtrsim 2$~keV) wavebands are available \citep{smp+07,gkk+11}. For each \swift\ GRB, $A_{V,GRB}$ is thus well measured. Note that this is the host galaxy visual extinction along the GRB line of sight, and thus does not necessarily reflect the galaxy-averaged visual extinction.

No optical or NIR afterglow was detected for GRB~000210 ($z=0.846$), although the derived deep afterglow upper limits ($R>23.5$ at 12.4~hrs after the GRB trigger) and relatively bright X-ray afterglow suggests that the optical/NIR afterglow was extinguished by significant amounts of dust along the line of sight \citep{pfg+02}. GRB~000418 was detected at optical and NIR wavelengths. However, the lack of an X-ray detection, and the relatively late and thus dim optical/NIR afterglow detection provided only a fairly crude measure of the visual extinction in the range $A_{V,GRB}=0.4-0.9$~mag \citep{ksm+00,bdf+01}. Nevertheless, $A_{V,GRB}=0.4$~mag would place GRB~000418 within the 25\% most dust-extinguished GRB afterglows at any redshift.

In the following subsections we provide a brief description of the host galaxy data available for our sample, all of which were taken once the afterglow had faded below the sensitivity limit of the instruments.

\begin{table}
\caption{\herschel\ PACS and SPIRE photometric measurements.}\label{tab:HerschPhot}
\begin{center}
\begin{tabular}{@{}lccccc}
\hline
& \multicolumn{5}{c}{Flux density (mJy)}\\
& \multicolumn{2}{c}{PACS} & \multicolumn{3}{c}{SPIRE} \\
\hline
GRB Host & 100$\mu$m & 160$\mu$m & 250$\mu$m & 350$\mu$m & 500$\mu$m \\
\hline\hline
000210 & $<5.1$ & $<11.1$ & $<24.0$ & $<36.4$ & $<38.6$ \\
000418 & $-$ & $-$ & $<22.5$ & $<35.8$ & $<33.4$ \\
070306 & $4.4^{+1.0}_{-1.0}$ & $6.2^{+1.2}_{-1.2}$ & $-$ & $-$ & $-$ \\
081109 & $<4.0$ & $<6.6$ & $<26.8$ & $<27.5$ & $<18.8$ \\
090926B & $<3.2$ & $<5.3$ & $<22.0$ & $<23.2$ & $<26.3$ \\
\hline
\end{tabular}
\end{center}
Notes: Upper limits are given at 3$\sigma$ confidence
\end{table}

\begin{figure*}
\centering
\includegraphics[width=1.0\textwidth]{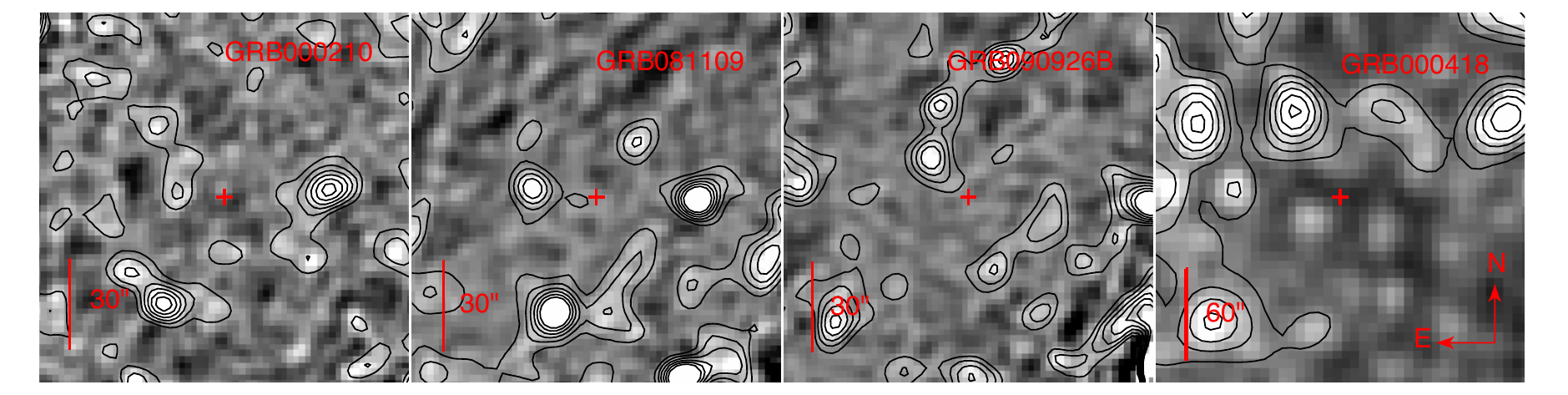}
\caption{\herschel/PACS 100$\mu$m images of GRBH~000210, GRBH~081109 and GRBH~090926B (first three panels), and \herschel/SPIRE 250$\mu$m image of GRBH~000418 (furthest right). The PACS and SPIRE images are 2\arcmin$\times $2\arcmin and 4\arcmin$\times $4\arcmin, respectively, centred on the GRB afterglow position (red cross), and have been smoothed using a Gaussian with $\sigma$ two times the pixel scale. The images are displayed with a linear greyscale ranging from $-1$~mJy~pix$^{-1}$ (black) to $+1$~mJy~pix$^{-1}$ (white), and contour levels from 1$\sigma$ to 6$\sigma$ are over plotted in back.}\label{fig:ULsample}
\end{figure*}

\begin{figure*}
\centering
\includegraphics[width=0.49\textwidth]{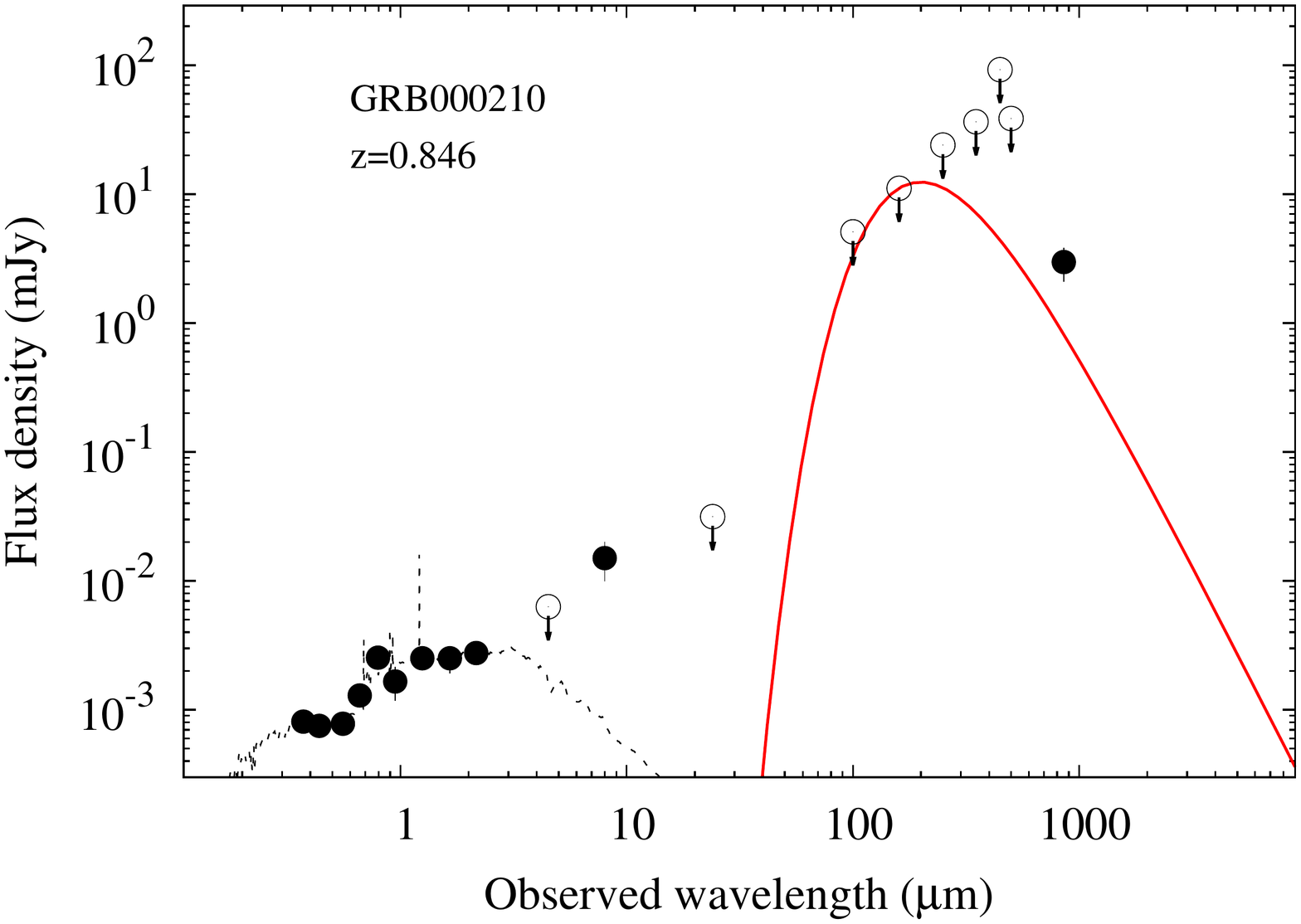}
\includegraphics[width=0.49\textwidth]{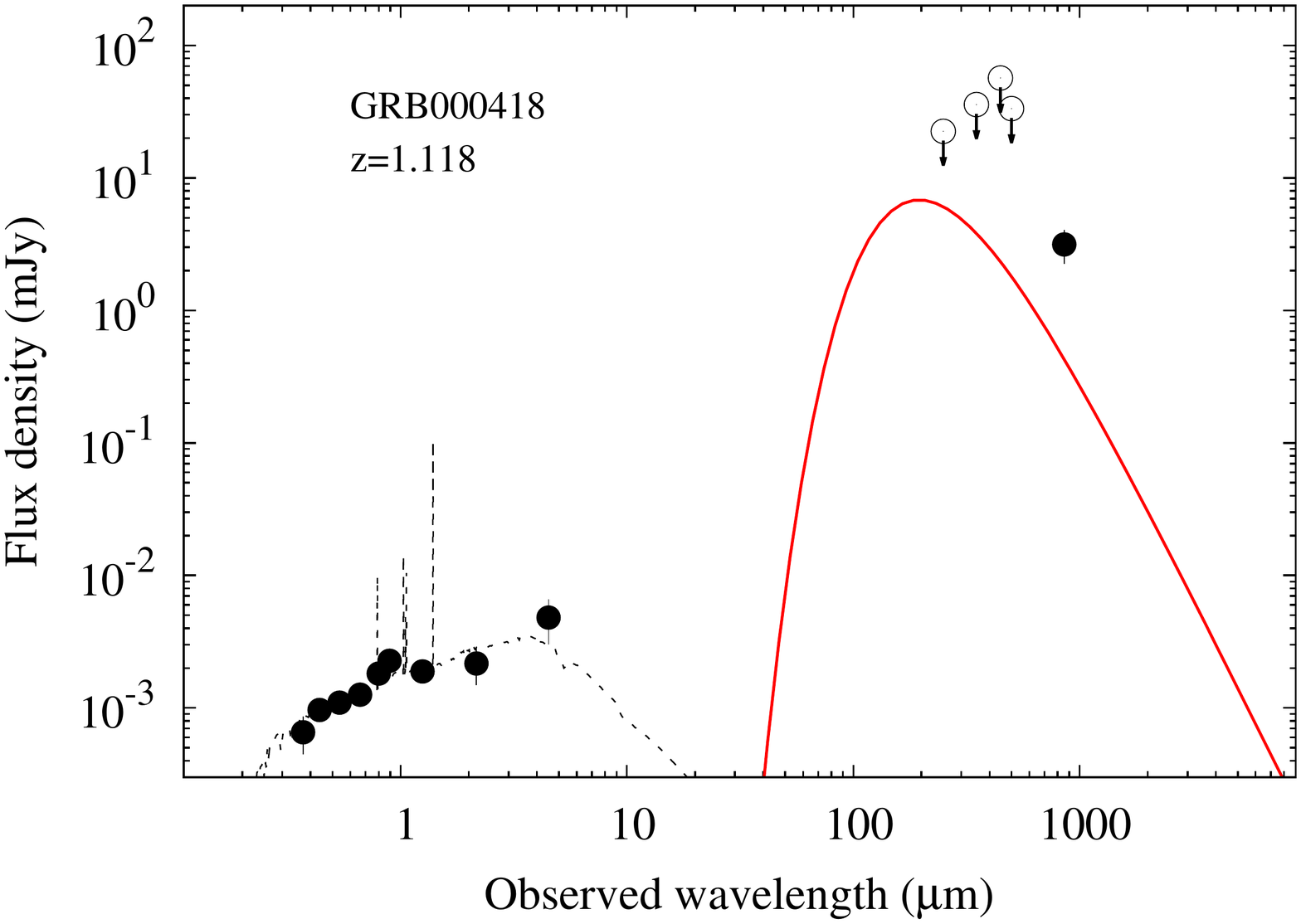}
\includegraphics[width=0.49\textwidth]{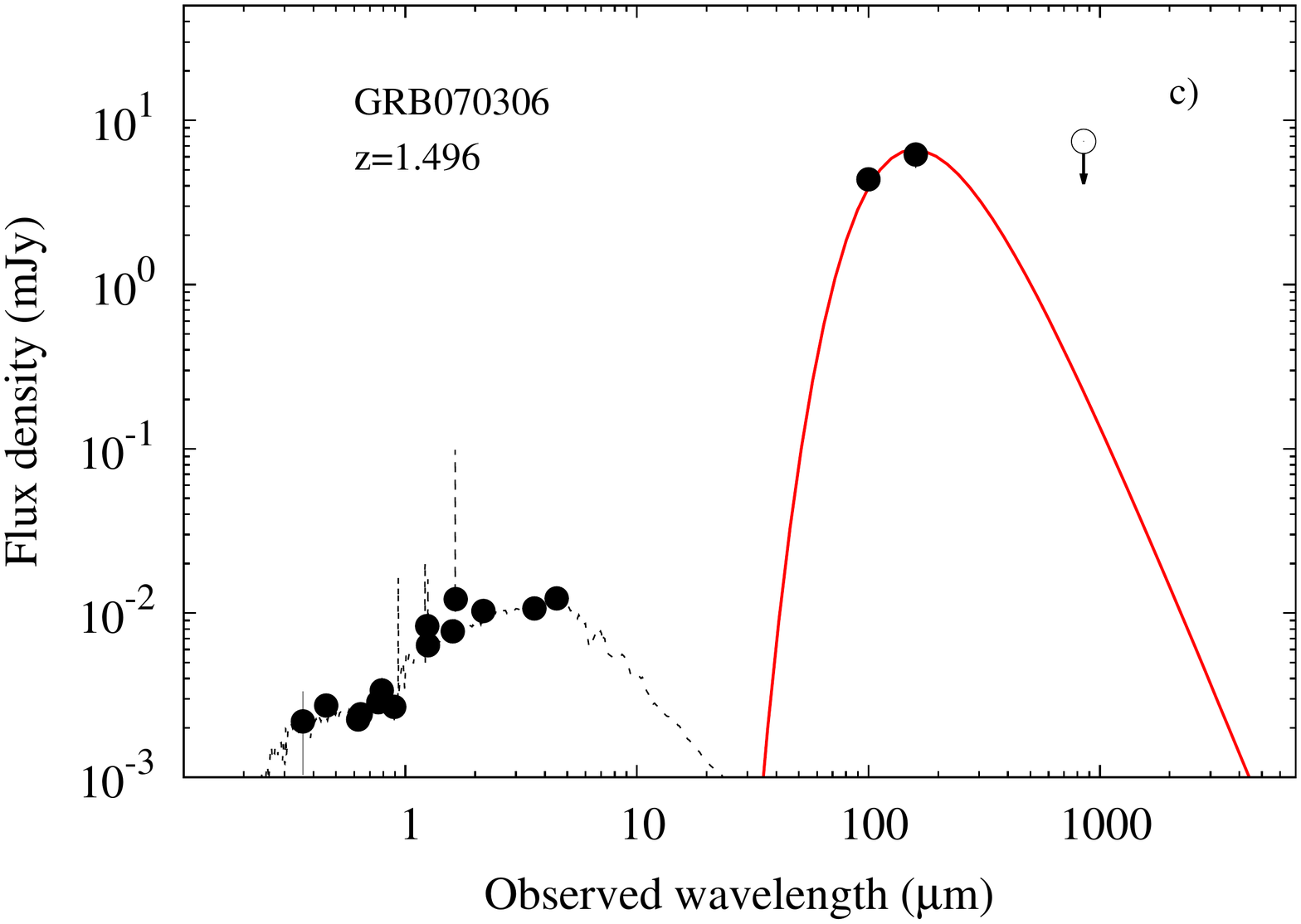}
\includegraphics[width=0.49\textwidth]{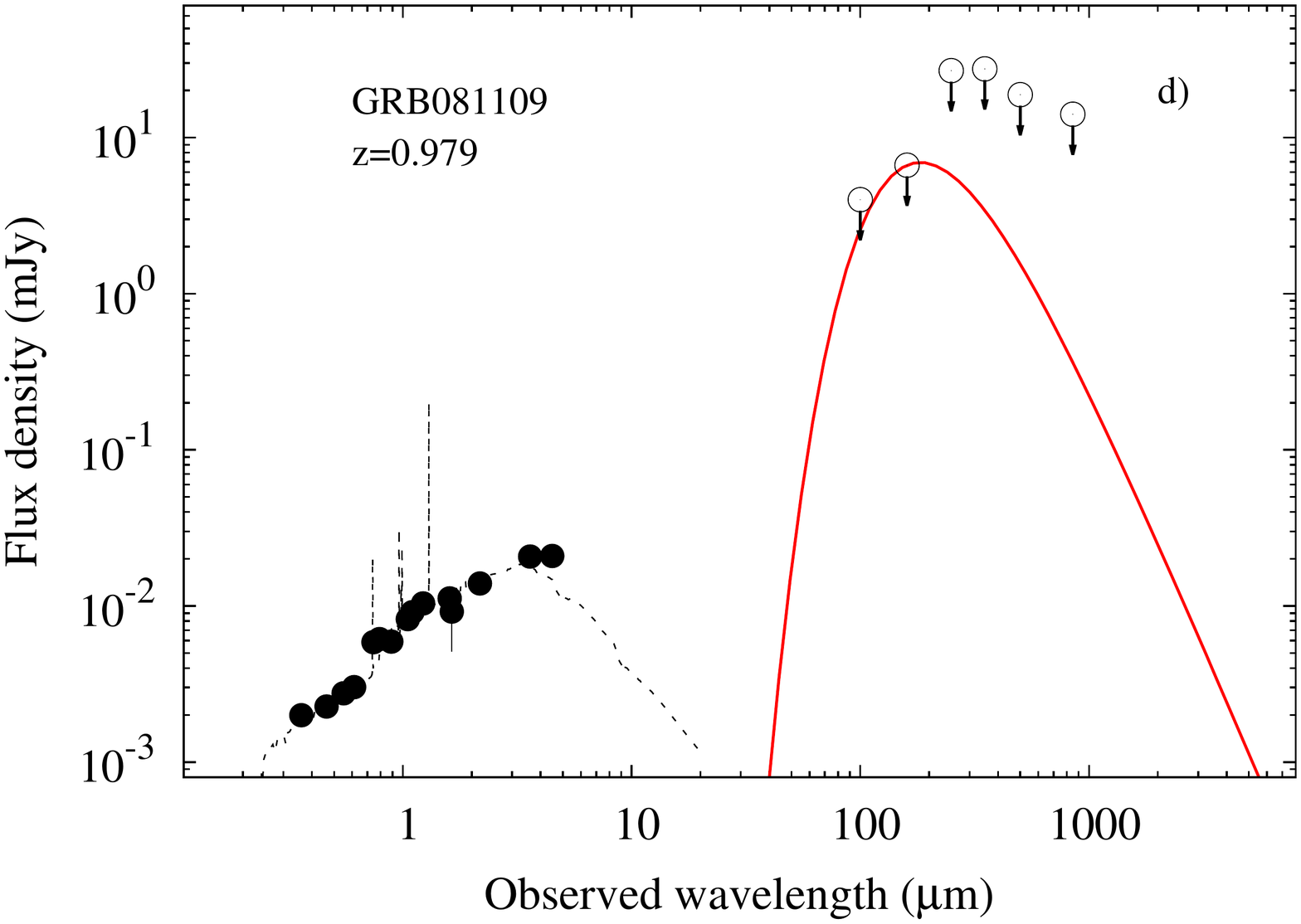}
\includegraphics[width=0.49\textwidth]{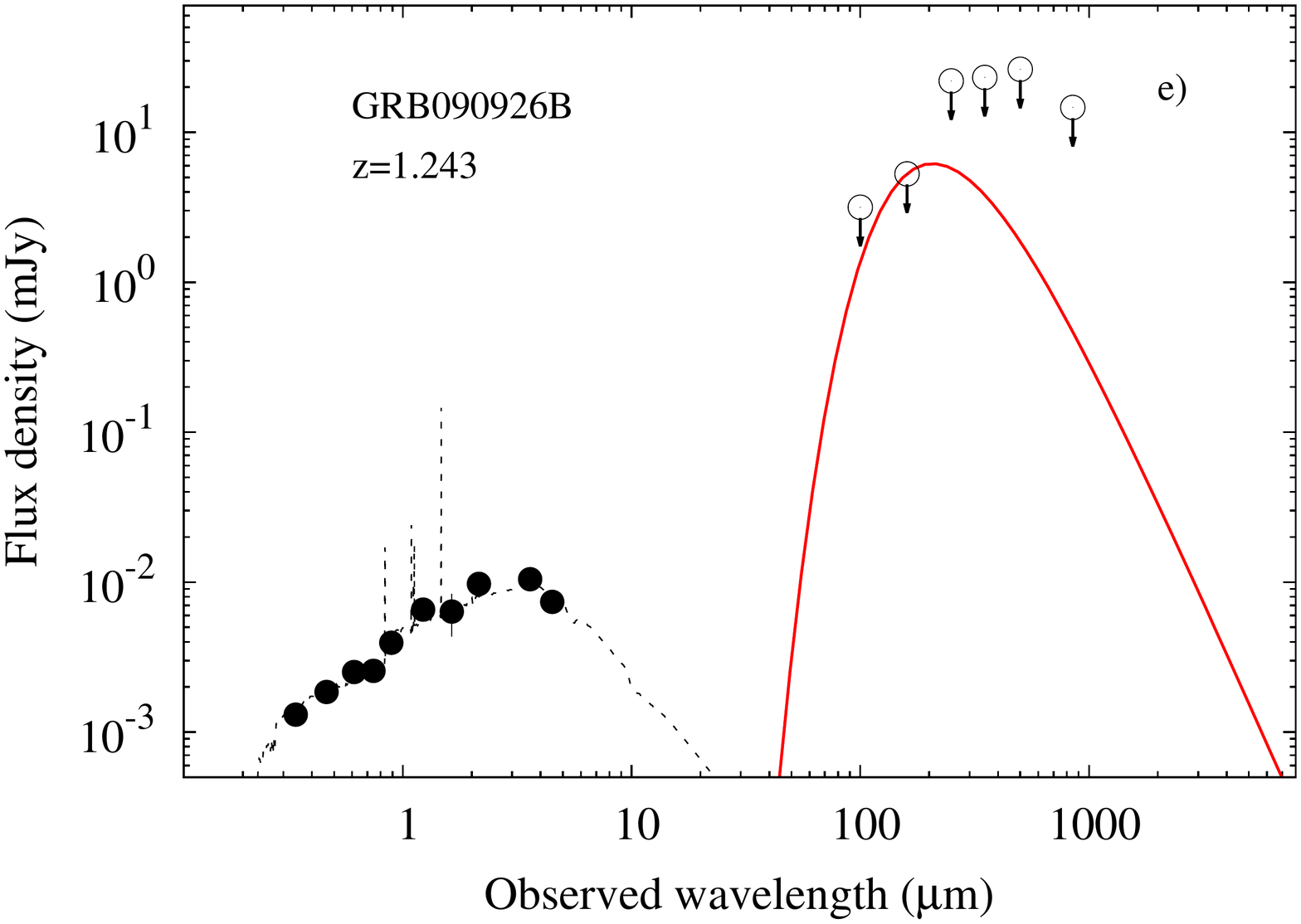}
\caption{GRB host galaxy optical through to submm spectral energy distributions for our sample of five GRBHs observed with \herschel. Detections are plotted as filled symbols, and open circles represent $3\sigma$ upper limits. The {\em lePHARE} best-fit template galaxy models fitted to the optical to mid-IR data are shown (dashed black line), as well as our modified blackbody fits to the FIR and submm data (red solid line) (see Table~\ref{tab:SEDfits} for details). In those cases where the host galaxy was not detected with \herschel, the modified blackbody fits were used to determine the upper limit on the corresponding dust mass and SFR.}\label{fig:SEDs}
\end{figure*}

\begin{figure*}
\centering
\includegraphics[width=0.49\textwidth]{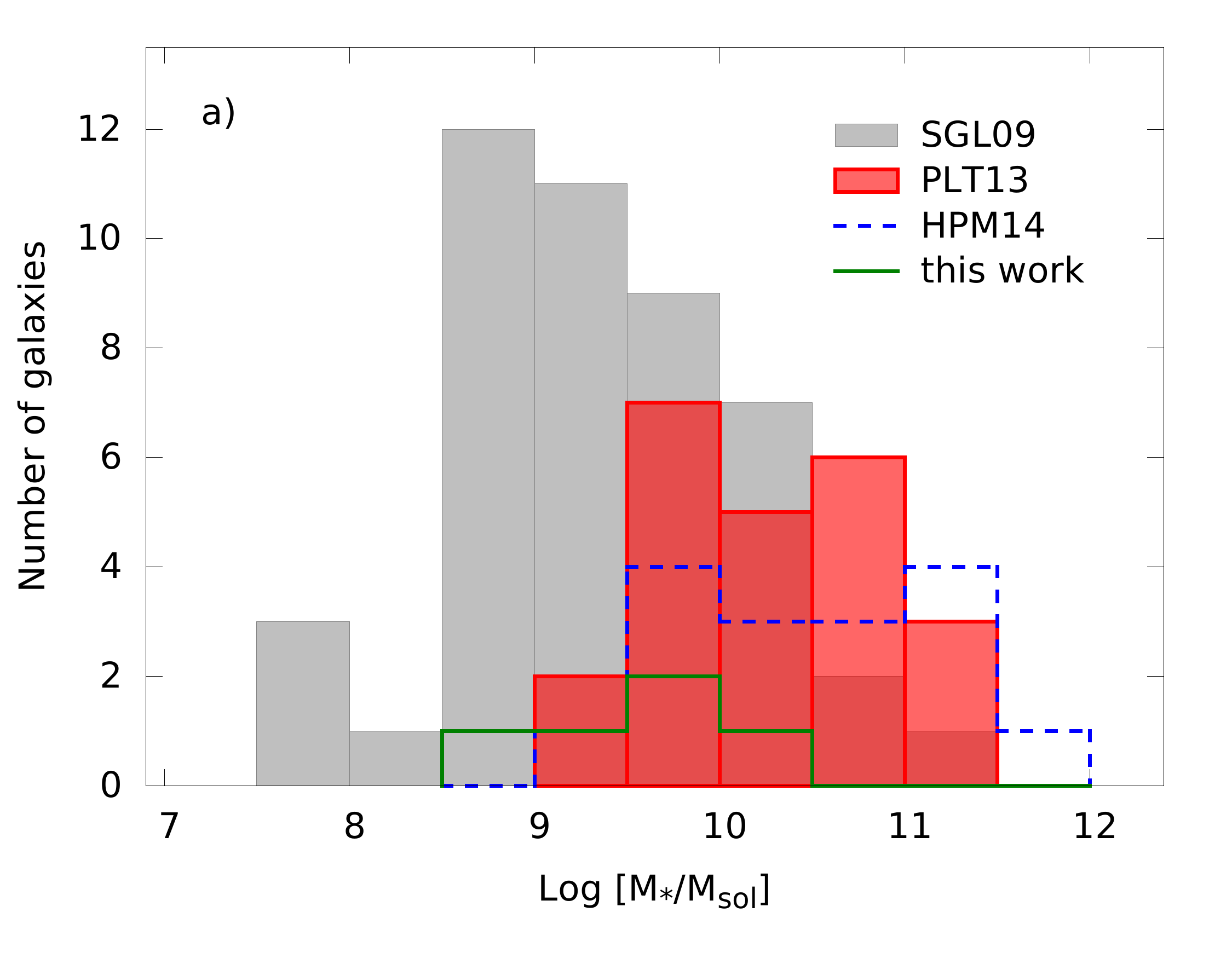}
\includegraphics[width=0.49\textwidth]{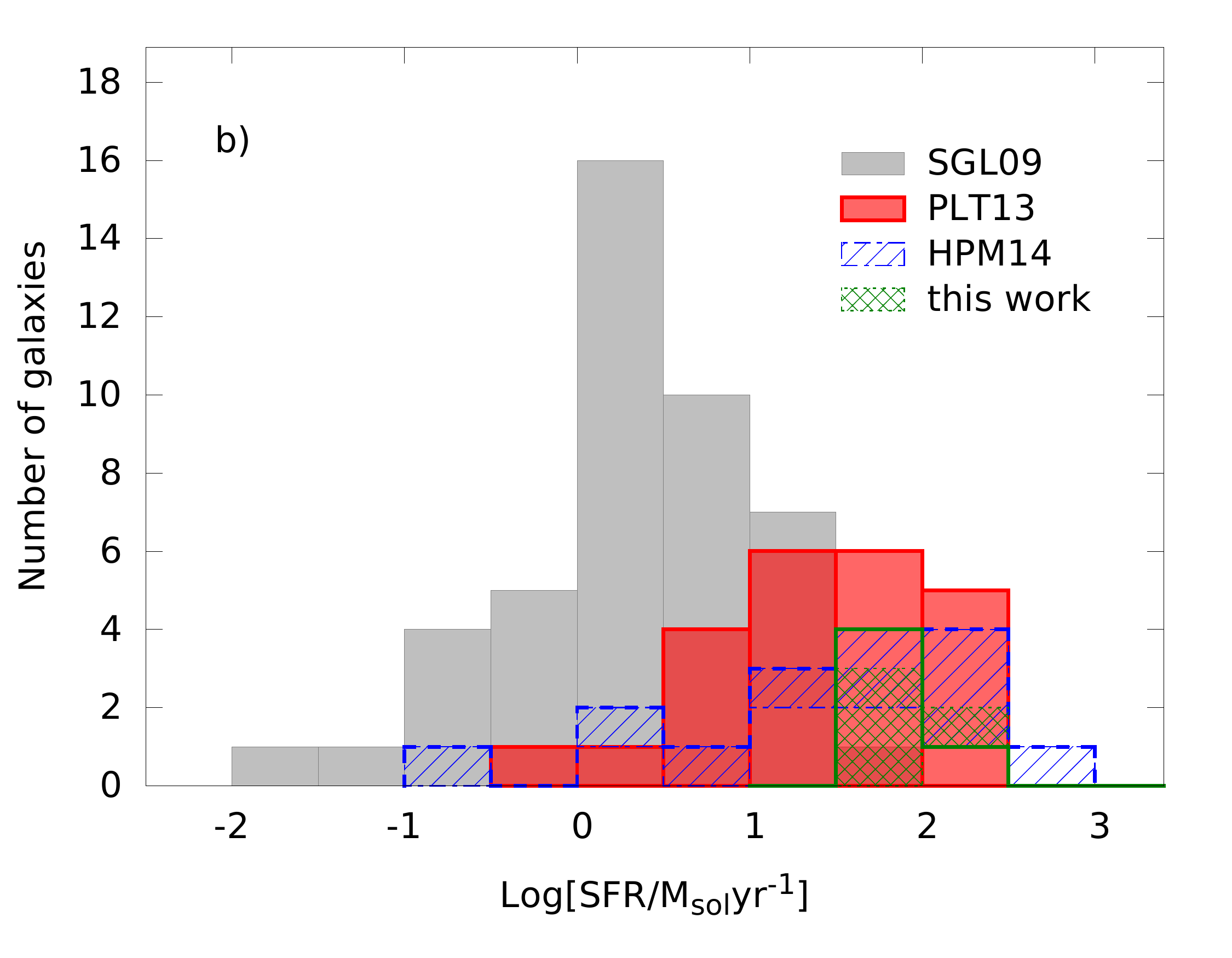}
\includegraphics[width=0.49\textwidth]{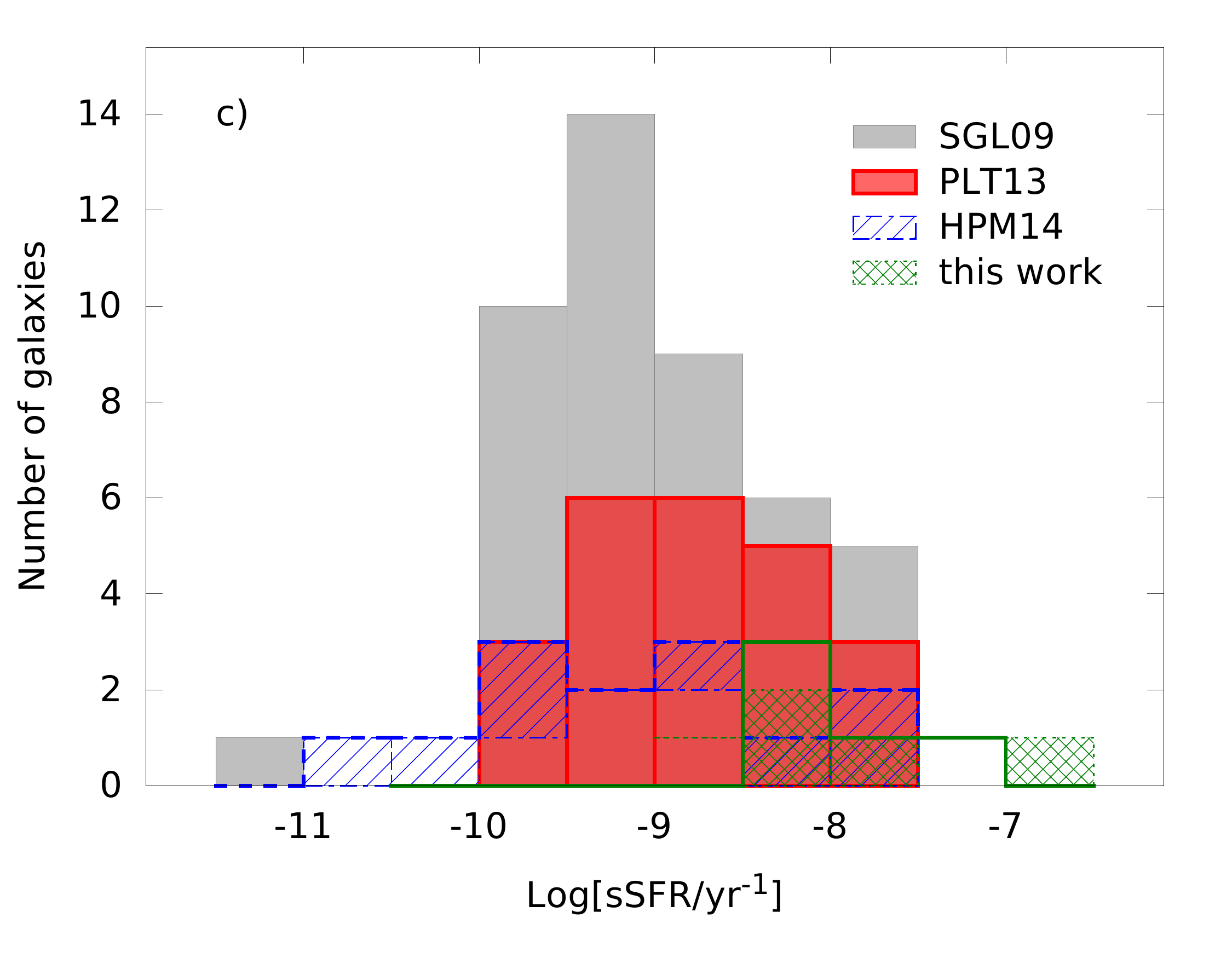}
\includegraphics[width=0.49\textwidth]{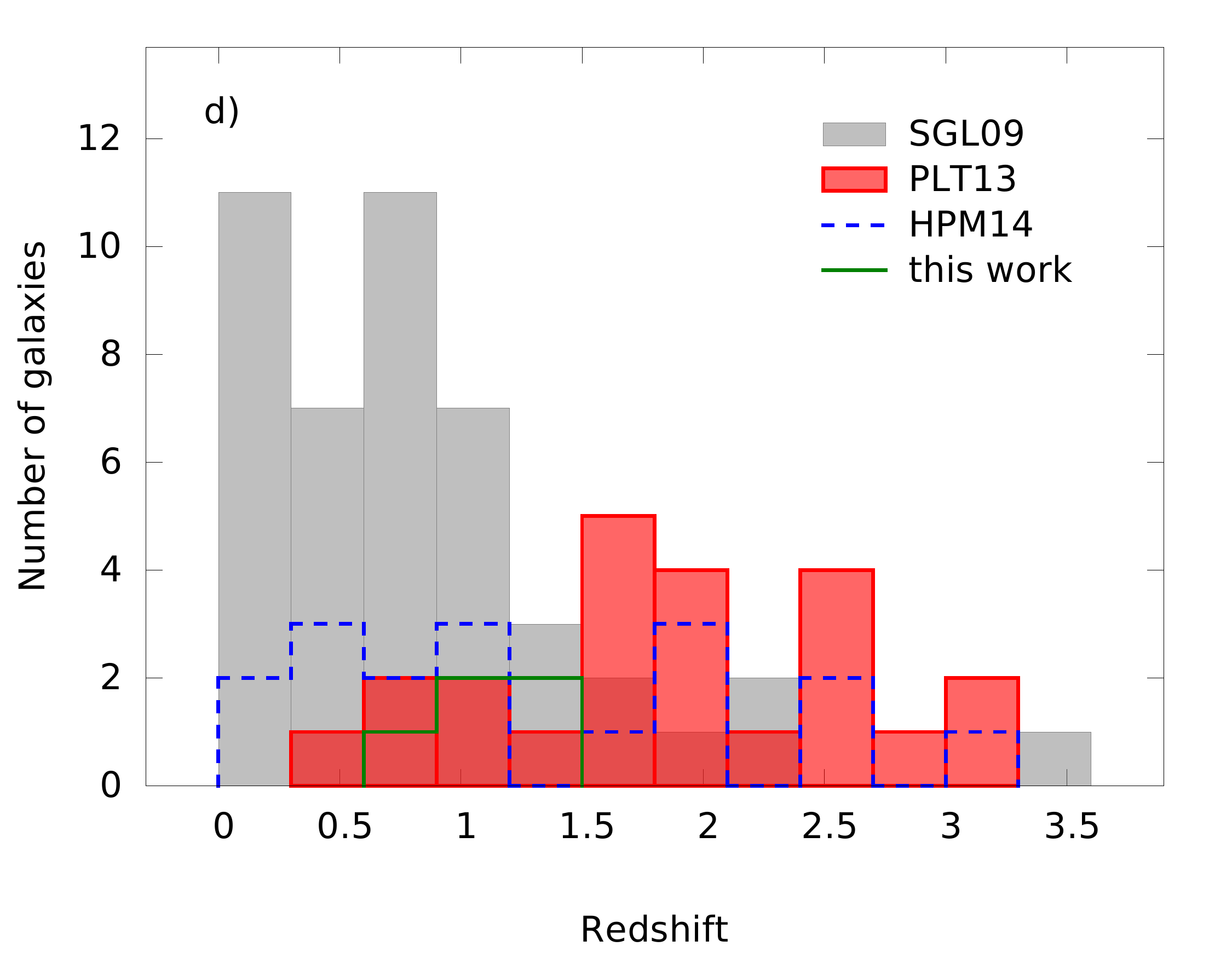}
\caption{Distribution of logarithmic $M_\ast$ (top left), SFR (top right) and sSFR (bottom left) as derived from SED analysis, and the redshift distribution (bottom right) from a number of GRB host galaxy samples. Outlined in solid green is the sample in this paper, in dashed blue is the sample from HPM14, the sample of host galaxies of heavily dust-extinguished GRBs ($A_{V,GRB}>1$~mag) from PLT13 is in filled red, and the filled grey histogram corresponds to the sample of hosts of GRBs with optically bright afterglows from SGL09. The samples from HPM14 and this work include SFR and sSFR upper limits, which are shown in hashed blue, and in a crisscross green pattern, respectively.}\label{fig:Hostprops}
\end{figure*}

\subsection{Host galaxy observations}
\subsubsection{Optical and near-infrared data}
\label{ssec:OptIRData}
All GRBHs in our sample have a large quantity of published optical and NIR data, as well as \spitzer\ data, which in the case of GRB~090926B have not been published, and we thus analysed the data ourselves. We list all optical and NIR data used in our SED analysis in Tables~A\ref{tab:OptPhot}-A\ref{tab:spitPhot}, together with corresponding references.

\subsubsection{Submillimetre data}
\label{ssec:submmData}
Prior to 2010, around 25 GRB host galaxies had been observed at 850$\mu$m with SCUBA \citep{bck+03,ttb+04}, of which only the hosts of GRB~000210, GRB~000418, and GRB~010222 were reported to have detections at the $3\sigma$ level. Since then, the submm emission thought to have come from the host galaxy of GRB~010222 has been put in doubt, likely originating in an unrelated, nearby source \citep{mhc+08}. We observed the remaining three host galaxies in our sample at 850$\mu$m with the LABOCA instrument on APEX over two consecutive semesters during MPG guaranteed time (PI: Greiner). The observations were taken between April and August in 2011, and amounted to an average on-source integration time of 1.5~hours per source. All submm flux measurements are listed in Table~A\ref{tab:submmPhot}.

\subsubsection{\herschel\ observations}
\label{ssec:HerschData}
Our sample of GRB host galaxies was observed with the {\em Herschel Space Observatory} at 100$\mu$m and 160$\mu$m using the PACS small-scan map mode (20\arcsec/s) \citep{pwg+10}, and all three SPIRE 250$\mu$m, 350$\mu$m and 500$\mu$m bands using the small-map mode \citep{gaa+10}. Our measurements reached typical 1$\sigma$ sensitivities of 1.0mJy, 2mJy, 8mJy, 10mJy and 20mJy at 100$\mu$m, 160$\mu$m, 250$\mu$m, 350$\mu$m and 500$\mu$m, respectively (included instrumental and confusion noise). The PACS observations of GRBH~000210 were shorter than for the rest of the sample, and thus reached 1$\sigma$ sensitivities of about 2mJy and 4mJy at 100$\mu$m and 160$\mu$m, respectively. In the case of the SPIRE observations the sensitivities were predominantly determined by the confusion limit, which is 2-3 times larger than the instrument point source sensitivity. The PACS and SPIRE beam FWHM sizes are 6.8\arcsec, 11.4\arcsec, 17.6\arcsec, 23.9\arcsec\ and 35.2\arcsec\ at 100$\mu$m, 160$\mu$m, 250$\mu$m, 350$\mu$m and 500$\mu$m, respectively. \herschel\ images for each GRBH in the bluest band observed are shown in Figs.~\ref{fig:DETsample} and \ref{fig:ULsample}.

Given the large uncertainty associated with predicting the FIR host galaxy flux from the UV through to mid-IR spectrum, we chose to initially observe our sample with SPIRE, in order to get good coverage of the thermal dust emission component. For the redshift range of our GRBH sample, we expected the dust emission to peak at $\sim 200\mu$m, observed frame. The three SPIRE bands just redward of this would thus have allowed relatively accurate determination of the shape of the thermal dust component. Depending on the flux that we measured within the SPIRE bands, we then re-adjusted the exposure times used for our PACS observations within the total observing time available for our programme. We did not detect any of the four GRB hosts initially observed with SPIRE (see Table~\ref{tab:obsLog}). In the case of GRBH~070306, which had a full-visibility window fairly late in the OT2 observing period, we therefore chose to only observe in PACS, which although provides coverage over a smaller wavelength range\footnotemark[2], is more sensitive than SPIRE. Finally, in the case of GRBH~000418, our broadband SED template fits to the optical/NIR data and SPIRE upper limits resulted in estimated flux densities that were below the sensitivity limit of PACS. We therefore did not observe GRBH~000418 with PACS, instead choosing to redistribute this time within our PACS observations. Details of the \herschel\ observations are listed in Table~\ref{tab:obsLog}.
\footnotetext[2]{Although the PACS photometer offers three bands (70$\mu$m, 100$\mu$m and 160$\mu$m), the 160$\mu$m filter can only be used with one of the other two filters simultaneously. For our observations we opted for the 100/160$\mu$m combination.}

\begin{table*}
\caption{GRB host galaxy properties taken from the literature and based on UV, optical, near- and mid-IR data, as well as the dust mass and temperature derived in this paper from modified blackbody fits to our \herschel\ data.\label{tab:SEDfits}}
\begin{center}
\begin{tabular}{@{}lcccccccc}
\hline
GRB & E(B-V)$^{(a)}$ & [O/H]$^{(b)}$ & SFR(H$\alpha$/O{\sc ii})$^{(c)}$ & SFR(FIR) & $\log[M_\ast]$ & A$_V^{(d)}$ & T$_d^{(e)}$ & $\log[M_d]^{(f)}$ \\
 Host & (mag) & & ($M_\odot$~yr$^{-1}$) & ($M_\odot$~yr$^{-1}$) & (M$_\odot$) & (mag) & (K) & (M$_\odot$) \\
\hline\hline
000210 & 0.029 & $8.3\pm 0.1$ & $1.1^{+0.1}_{-0.1}$ & $<37$ & $9.23^{+0.12}_{-0.08}$ & $0.0$ & $30$ & $<8.6$\\
 & & $8.7\pm 0.1$ & & & & & &\\
000418 & 0.033 & $8.1\pm 0.2$ & $26^{+1.2}_{-1.4}$ & $<40$ & $8.87^{+0.15}_{-0.10}$ & $1.3$ & 40 & $<8.3$\\
 & & $8.8\pm 0.2$ & & & & & &\\
070306 & 0.024 & $8.4\pm 0.1$ & $61^{+10}_{-10}$ & $101\pm21$ & $10.34^{+0.09}_{-0.29}$ & $0.4$ & $51\pm 0.2$ & $7.9\pm 0.3$\\
081109 & 0.017 & $8.8\pm 0.1$ & $24^{+14}_{-9}$ & $<32$ & $9.93^{+0.06}_{-0.04}$ & $1.3$ & 35 & $<8.5$\\
090926B & 0.020 & $8.2\pm 0.2$ & $12^{+12}_{-6}$ & $<45$ & $9.76^{+0.07}_{-0.05}$ & $1.0$ & 35 & $<8.7$\\
\hline
\end{tabular}
\end{center}
{\tiny $^{(a)}$ Galactic dust reddening in magnitudes from the map of \citet{sf11}\\
$^{(b)}$ Metallicities in the case of GRBH~000210, 000418 and 070306 \citep{pir14, ver14} were derived using the $R_{23}$ ratio and applying the calibration from \citet{kk04}. The N~{\sc ii} emission line required to distinguish between the $R_{23}$ lower and upper branches was not detected in the spectra of GRBH~000210 and 000418, and we therefore report both solutions. Metallicities for GRBH~081109 and 090926B \citep{kru+14} were derived using the methods from \citet{nmm06}. For a detailed description on the differences between the individual strong line diagnostics, see \citet{ke08}.\\
$^{(c)}$ SFRs derived from the O~{\sc ii} line for GRBH~000418 \citep{pir14}, and from the H$\alpha$ line for the other four GRBHs \citep{pir14,kru+14,ver14}, based on the formulation described in \citet{ken98}, but converted to a Chabrier IMF \citep{cha03}. All have been corrected for host galaxy dust extinction derived from the Balmer decrement, with the exception of GRBH~000418, for which we used the average visual extinction given in column 7, and assuming the Calzetti dust attenuation law \citep{cab+00}.\\
$^{(d)}$ The quoted value corresponds to the $A_V$ of the best-fit galaxy template.\\
$^{(e)}$ Temperature of the MBB scaled to the data, which is fixed, apart from in the case of GRBH~070306, where it was left as a free parameter in the SED fit.\\
$^{(f)}$ 3$\sigma$ upper limit or best-fit dust-mass resulting from a blackbody scaled to the data with emissivity index $\beta=1.5$ and temperature as given in column 8.}
\end{table*}

\subsubsection{\herschel\ data reduction}
\label{sssec:Hreduct}
Data reduction in PACS and SPIRE was performed using the \herschel\ Interactive Processing Environment (HIPE v10.0.0) \citep{ott10}. The PACS data were reduced using the `deep survey point-source' script within HIPE. We used pixel sizes of 2\arcsec\ and 3\arcsec\ for PACS 100$\mu$m and 160$\mu$m, and 6\arcsec, 10\arcsec\ and 14\arcsec\ for SPIRE 200$\mu$m, 350$\mu$m and 500$\mu$m, respectively. The individual PACS scans were processed with a high pass filter to remove $1/f$ noise and thermal drifts in the PACS bolometers. We used a running box median filter with a half-width of 31 frames (62\arcsec) at 100$\mu$m and 51 frames (102\arcsec) at 160$\mu$m. This choice of the high-pass filter radius allows us to optimise the 1/f noise subtraction, thus reducing the final map noise without degrading the PACS PSF. We initially stacked the cross-scans to create a mask from this deeper image, and the individual cross-scans were then re-reduced and stacked, this time using our newly created mask file to mask out the bright regions. Finally, we corrected the astrometry in our final, stacked, images using the aspect solution from the bluest \spitzer\ Infrared Array Camera observations available for each field. These corrections were typically of the order of 2\arcsec\ at 100$\mu$m at 3\arcsec\ at 160$\mu$m.

Fluxes were measured at the source location using the HIPE internal aperture photometry routines and checked with IDL-based procedures. For the SPIRE images, aperture radii of 22\arcsec, 30\arcsec\ and 42\arcsec\ were used at 250$\mu$m, 350$\mu$m and 500$\mu$m, respectively, in line with SPIRE calibration guide lines \citep{pln+13}. For the PACS stacked images, we used radii of 5\arcsec\ and 7.5\arcsec\ at 100$\mu$m and 160$\mu$m, respectively, in order to gain the highest signal-to-noise ratio for a point source \citep{per+10}, applying a respective aperture correction of 1.9 and 2.0 \citep{bmn+13}. Colour corrections are of the order of unity, and we therefore neglect them \citep{pwg+10}.

In the case of GRBH~070306, there is a source 10\arcsec\ north-east of the GRB host galaxy. In order to disentangle the flux emission from the host galaxy of GRB~070306 and the nearby source, we used the GALFIT package \citep{phi+02} to simultaneously measure the flux from the two sources based on the measured \herschel\ PSF. At 160$\mu$m the FWHM beam size (11.4\arcsec) does not allow the GRB host galaxy and neighbouring source to be resolved, and we therefore fixed the position of the two sources to the 100$\mu$m best-fit positions found with GALFIT. The 100$\mu$m source positions were within 0.3\arcsec\ and 1.6\arcsec\ of the GRBH and nearby source centroid positions, respectively, in the 4.5$\mu$m \spitzer\ image.

In all cases, the rms sky noise was calculated from the standard deviation of 100 circular apertures of the same size as the source extraction region randomly placed around the source position, within high-coverage regions of the image. This provides an accurate measure of the sky background and includes correlated noise \citep[see section 4.2.2 in][]{bmn+13}. Our SPIRE and PACS flux measurements are listed in Table~\ref{tab:HerschPhot}.

\section{Data analysis}
\label{sec:rslt}
Of the five GRB host galaxies in our sample, only GRBH~070306 was detected at 100$\mu$m and 160$\mu$m. No SPIRE observations of this host galaxy were taken (see Fig.~\ref{fig:SEDs}). Given that by selection, all the host galaxies in our sample showed evidence of dust along the GRB line of sight, the lack of a \herschel\ detection in four out of five cases may imply that the dust mass within these galaxies is generally low, and concentrated in a few discrete locations within the galaxy. In the following section we describe our analysis and provide our best-fit results from our SED fits.

\subsection{Spectral energy distributions}
\label{ssec:seds}
\footnotetext[3]{Only GRBH~000210 was observed within this wavelength range at 24$\mu$m with \spitzer.}
\footnotetext[4]{Two detections of GRBH~070306 with PACS at 100$\mu$m and 160$\mu$m, and two tenuous detections of GRBH~000210 and GRBH~000418 at 850$\mu$m with SCUBA.}
For each GRB host galaxy in our sample we combined all optical through to submm flux density measurements to create a broadband SED. Most of our sample SEDs have no data between 10$\mu$m and 100$\mu$m \footnotemark[3], and there are few detections redward of 25$\mu$m \footnotemark[4] (see Fig.~\ref{fig:SEDs}). We therefore chose to place constraints on the host galaxy dust properties by fitting just the FIR and submm data separately with a modified blackbody (MBB). The emissivity index was set to $\beta=1.5$ and the grain absorption cross section per unit mass was set to $\kappa_{\rm abs}=3.4$~cm$^{2}$~g$^{-1}$ at 250$\mu$m rest frame \citep{bia13}. These values are consistent with the thermal emission measured in other star forming galaxies \citep[i.e. $\beta=1.5-2$][]{de01,dae+12,cas12}. In the case of GRBH~070306, the dust temperature was constrained by the two PACS detections, and thus this parameter was left free to vary. For the remaining four GRBHs, the temperature was fixed to a typical value of 35K \citep{sed+11,msl+12,rmg+13,szs+14}, although in the case of GRBH~000210 and GRBH~000418, the SCUBA detections provided further constraint on the dust temperature, and we found that temperatures of 30K and 40K, respectively were more compatible with the data (see sections~\ref{ssec:grb000210} and \ref{ssec:grb000418} for details).

From our SED fits, we determined the upper limit on the dust mass allowed by our data, which was mostly constrained by the bluest PACS data at 100$\mu$m, and we used the total luminosity (8-1000$\mu$m) measured from our best-fit MBB model to estimate the FIR SFR using the prescription from \citet{ken98}. In all cases, the FIR-derived SFR or upper-limits are higher than the optically derived SFRs by at least 30\%, suggesting that, at least in the case of GRBH~070306, there is some obscured star formation (see Table~\ref{tab:SEDfits}).

We repeated our fits for an emissivity index $\beta=2$, and found that this has a weak effect on our total IR luminosity, but it typically decreases the dust mass by 0.3dex. A change in the blackbody temperature has a much stronger effect, with the dust mass varying over two orders of magnitude for dust temperatures in the range 20-50K, and the SFR varying by $\sim 30$\%. This uncertainty is most relevant for GRBH~0811109 and GRBH~090926B, for which there was no detected FIR or submm emission with which to constrain the dust temperature. However, both these host galaxies have relatively high un-obscured star formation rates (see Table~\ref{tab:SEDfits}), and thus would be expected to have relatively high dust temperatures ($>35$K) \citep{hpm+14,szs+14}. The dust mass is inversely related to the dust temperature, and we therefore consider our dust mass upper limits for both GRBH~081109 and GRBH~090926B to be fairly conservative in the sense that the true dust mass is likely to be discernibly lower than our stated limit. A more important source of error is in our assumption of a single-temperature MBB, which does not take into account colder dust that can make up 50\% of the total dust mass \citep[e.g.][]{dae+12,mdb+12,mrh+13}. For those galaxies without a submm detection, for which the emission from a cold-dust component cannot be constrained, we increased our dust mass errors to 0.3dex to account for the uncertainty in the dust temperature. The broadband SEDs for each GRB host in our sample are shown in Fig.~\ref{fig:SEDs} together with the MBB model.

Data below $25\mu$m were fitted with the spectral template fitting package {\em LePHARE} \citep{acm+99,iam+06}, which is a population-synthesis-based fitting procedure. We used the \citet{bc03} galaxy templates, which include emission lines and prescribed reddening and parameters therein, and we assumed a \citet{cha03} IMF.

\subsection{Comparison with the literature}
\subsubsection{Stellar mass and SFR}
After correcting for differences in the assumed IMF in this work and in other publications (i.e. \citet{mhc+08} assumed a \citet{sal55} IMF and \citet{sgl09} (SGL09 from here on) assumed a \citet{bg03} IMF), our stellar masses are in agreement (at 1$\sigma$) with other reported values (i.e. SGL09; Kr{\"u}hler et al., 2011; PLT13; HPM14), with the exception of the higher $M_\ast$ found by \citet{mhc+08} for GRBHs~000210 and 000418. After converting to a \citet{cha03} IMF, the $M_\ast$ in \citet{mhc+08} is a factor of 3 and 14 higher than ours, respectively, which may be the result of differences in the star formation histories and stellar populations assumed in the modelling \citep[][see section~\ref{sec:HostProps}]{mdc+12}. The SED fitting software package, GRASIL \citep{sgb+98}, which accounts for the stellar mass from a starburst as well as from continuous star formation, can increase the best-fit stellar mass by up to 0.4dex (relative to optical/NIR-only SED fits). The inclusion of FIR data used in \citet{mhc+08} may also increase $M_\ast$ if the best-fit templates include a fully dust-obscured stellar population \citep[e.g.][]{lfv+13}.

There is greater scatter amongst the optically derived SFRs. This is related to the uncertainty in the dust correction, which can vary by a factor of a few. Our SFR(FIR) for GRBH~070306 is consistent with HPM14, but our values for GRBHs~000210 and 000418 differ from the results in \citet{mhc+08} by a factor of 4 and 2, respectively. We comment on the possible reasons for this below.

\subsubsection{Dust mass and temperature}
\subsubsection*{GRBH~000210}
\label{ssec:grb000210}
This galaxy had a reported 850$\mu$m 3$\sigma$ detection \citep{bck+03,ttb+04}, with S$_{850\mu{\rm m}} = 3.0\pm 0.9$mJy. A MBB with T=35K scaled to our \herschel\ limits underestimates the thermal emission at 850$\mu$m by one to two orders of magnitude for an emissivity index $\beta=1.5-2$. A MBB with a colder dust temperature of T=30K is in closer agreement to the \herschel\ and SCUBA observations, but only just at the 3$\sigma$ level (Fig.~\ref{fig:SEDs}a, red line). Although a MBB peaking at even colder dust temperatures could theoretically be scaled up to the SCUBA flux measurement without exceeding the \herschel\ upper limits, the resulting dust mass would lead to unfeasibly high dust-to-stellar mass ratios. Some SMGs can have dust-to-stellar mass ratios as high as 0.3 \citep{mhw10}, but these are extreme cases and more typical values are closer to $\sim 0.1$. A dust temperature of T=30K and dust mass $\log[M_d/M_\odot]=8.6$ therefore provides the greatest consistency with the data.

\citet{mhc+08} found the host galaxy SED to be best-fit by a higher effective temperature of 45~K. However, these fits lacked our \herschel\ data, which place an upper limit on the amplitude of the thermal dust emission. In our model, the smaller normalisation of a MBB imposed by the \herschel\ upper limits (and thus lower 850$\mu$m emission) is counter-acted by a decrease in the dust temperature. The \herschel\ upper limits similarly result in a lower SFR(FIR) than in \citet{mhc+08}. The marginal consistency between our model and the submm flux measurement places the SCUBA detection somewhat in doubt, and in table~\ref{tab:SEDfits} we thus give this dust mass as an upper limit.

\subsubsection*{GRBH~000418}
\label{ssec:grb000418}
This galaxy was also detected at the 3$\sigma$ level with SCUBA, with a flux density S$_{850\mu{\rm m}}=3.2\pm 0.9$mJy \citep{bck+03,ttb+04}. The dust mass upper limit derived from a T=35K MBB fit to our \herschel\ data is only just consistent with the SCUBA detection at the 3$\sigma$ level. However, our best-fit dust mass of $\log[Md/M_\odot]=8.9$ is almost the same as the galaxy stellar mass, which as explained above, becomes difficult to explain physically. Applying a dust-to-gas ratio of 0.1 gives a considerably lower dust mass of $\log[Md/M_\odot]=8.3$. It is possible to remain within this dust mass upper limit and reproduce the SCUBA flux measurement (albeit only just within the 3$\sigma$ limit) by increasing the temperature of the warm component to 40K. When scaled to the PACS 100$\mu$m upper limit, a temperature any higher than 40K would fail to reproduce the SCUBA flux measurement.

The effective temperature fitted by \citet{mhc+08} was 50K, and their SFR(FIR) was also higher than our upper limit, and this is similarly related to the additional data coverage blueward of 450$\mu$m provided by our \herschel\ PACS data. For a similar reason as in the case of GRBH~000210, we report the dust mass of $\log[Md/M_\odot]=8.3$ as an upper limit.

\subsubsection*{GRBH~070306}
\label{ssec:grb070306}
This is the only host galaxy from our sample that was detected with \herschel. Assuming a single MBB component, the best-fit dust temperature and mass are T=$51.2\pm 0.1$K and $\log[M_d/M_\odot]=7.9\pm 0.3$.

This GRB host galaxy was also included in the sample of galaxies studied in HPM14, in which the optical through to radio host galaxy data were fitted simultaneously using the software package GRASIL. This treats the stellar light absorbed by dust, and the re-emitted dust emission at FIR and submm wavelengths in a self-consistent way, and includes emission from dust grains with a continuous distribution of temperatures, in addition to emission from mid-IR wavelengths from polycyclic aromatic hydrocarbons (PAHs). In addition to the different SED modelling used in this paper and in HPM14, the \herschel\ PACS photometry for GRBH~070306 also differed, in particular at 160~$\mu$m, although still consistent at the 2$\sigma$ level. The reason for this difference is most likely related to the methods used to remove the contamination from a nearby source (see section~\ref{sssec:Hreduct}). Whereas HPM14 applied aperture photometry within HIPE, we used the GALFIT software, as described in section~\ref{sssec:Hreduct}.

Despite the (small) differences in photometry, and the much simpler approach that we use in this paper to constrain the host galaxy dust properties, our best-fit dust mass ($\log[M_d/M_\odot]=7.9\pm 0.3$) is consistent within 1$\sigma$ with the GRASIL best-fit value of $8.3\pm 0.3$.

\subsubsection*{GRBH~081109}
\label{ssec:grb081109}
In a campaign led by our group to follow-up the host galaxies of significantly dust-extinguished GRBs ($A_{V,GRB}>1$~mag) with the LABOCA instrument on APEX, this galaxy was detected at the 3$\sigma$ level, with a flux density $S_{870\mu m}=18.0\pm 4.7$mJy. However, the non-detection of the host galaxy in both \herschel\ PACS and SPIRE bands would imply an unfeasibly low dust temperature ${\rm T}<10$K, indicating that the APEX detection was due to a spurious source or blending. Within the PACS and two bluest SPIRE images, there are three resolved sources within a 30\arcsec\ region around the GRB host position, all of which could have contributed to the flux density measured within the LABOCA 19\arcsec\ beam. We therefore conclude that the LABOCA detection was likely spurious or contaminated by an unrelated source, and determine a 3$\sigma$ upper limit on the host galaxy submm flux density of 14mJy. Assuming an average dust temperature of 35K, the PACS upper limits then constrain the dust mass to $\log[M_d/M_\odot]<8.5$.

\subsubsection*{GRBH~090926B}
\label{ssec:grb090926B}
This galaxy was undetected in all PACS and SPIRE bands, as well as at 870$\mu$m with LABOCA, with a $3\sigma$ upper limit of S$_{870\mu m}<15$mJy. The greatest constraint to a MBB with temperature T=35K is provided by the 160$\mu$m PACS upper limit, yielding a dust-mass upper limit of $\log[M_d/M_\odot]<8.7$.

\section{Summary of host galaxy properties}
\label{sec:HostProps}
Our sample was selected from the GRB optical afterglow properties, based on the expectation that dusty GRB sightlines are indicative of a host galaxy with a high dust mass. In order to see how these selection criteria affect the overall galaxy properties of the sample, here we briefly summarise some of the characteristic properties of our host galaxy sample (see Table~\ref{tab:SEDfits}), and compare these to other GRB host galaxy samples.

In Fig.~\ref{fig:Hostprops} we show the logarithmic distribution of $M_\ast$, SFR and sSFR for a number of GRB host samples, as well as the redshift distribution. Our sample is outlined by the solid green line. In filled grey we show the distribution of properties for the host galaxy sample of SGL09, which is made up of 46 galaxies that were selected on the basis of optical and NIR detections, and thus are predominantly the hosts of GRBs with optically bright afterglows (i.e. small $A_{V,GRB}$ at $z<1.5$). In contrast to this, we also show the sample from PLT13 (filled red), which is made up of 23 host galaxies of heavily extinguished GRBs ($A_{V,GRB}>1$~mag), and the sample of \herschel\ observed hosts from HPM14 (dashed blue outline), which was predominantly selected on the basis of two or more host galaxy \spitzer\ detections ($\gtrsim 10\mu$Jy at 3.6$\mu$m and 4.5$\mu$m), with a preference for hosts of GRBs with $\beta_{OX}<0.5$.

Stellar masses and SFRs were derived from SED fits to the UV through to near- or mid-IR data in the SGL09 and PLT13 samples, and in the case of HPM14, FIR \herschel\ data were also included. In the case of HPM14 and our sample, SFRs were determined from the IR luminosity, which in most cases are thus upper limits. This is indicated in Fig~\ref{fig:Hostprops}a and b with the blue hashed, and green crisscross pattern for the HPM14 and our samples, respectively.

Differences in the stellar population models and star formation history parameterisations used in the SED fits, as well as in the assumed dust attenuation, can introduce systematic differences in M$_\ast$ of up to 0.2-0.3dex \citep{pbl+07,kit+09,wfc+09,isl+10,mlb+13}. In particular, SGL09 used an IMF from \citet{bg03}, which gives a total stellar mass that is 0.2-0.3dex higher than when assuming a \citet{cha03} IMF, as was the case in the modelling of PLT13 and HPM14. Already without correcting for this, the stellar masses in SGL09 appear systematically lower than when compared to the other samples, and a Kolmogorov-Smirnoff (KS) test gives a 99.999\% probability that the PLT13 and HPM14 samples come from a different parent population to the SGL09 sample.

Systematic differences in the SFR arise predominantly from the uncertainty in dust attenuation. The SFRs in SGL09 were derived from UV and optical photometry, thus making the level of dust-obscured star formation highly uncertain. The inclusion of \herschel\ observations in our sample and in HPM14 provide a better handle on the total SFR, and thus the SFRs in these two samples are typically higher than the SFRs in SGL09. The SFR remains uncertain for those GRBHs in HPM14 that were not detected with \herschel, and in these cases we therefore use the optically derived (dust corrected) SFRs when comparing the HPM14 and SGL09 sample SFRs. We find that the two SFR distributions are inconsistent with coming from the same parent population, with a null-hypothesis probability P=0.01, and P only increases to 0.02 when we correct the SFRs to the same IMF. The high SFRs in the PLT13 sample are more likely related to the selection criteria rather than greater sensitivity to the total SFR (host galaxies with $A_{V,GRB}>1$~mag tend to be more massive and actively star forming). The sSFR in PLT13 and HPM14 are nevertheless consistent with the sSFR in SGL09 (P=0.2 and 0.5, respectively).

For the majority of the samples considered here, it is unlikely that cosmic evolution is responsible for the differences observed in the M$_\ast$ and SFR, since the redshift distributions are fairly consistent between samples. Only the PLT13 sample shows a very different redshift distribution, with a mean redshift of $\langle z\rangle=1.8$, compared to mean redshifts of $\langle z\rangle=$0.8, 1.1 and 1.1 for the SGL09, HPM14 and our sample.

\begin{figure}
\centering
\includegraphics[width=0.55\textwidth]{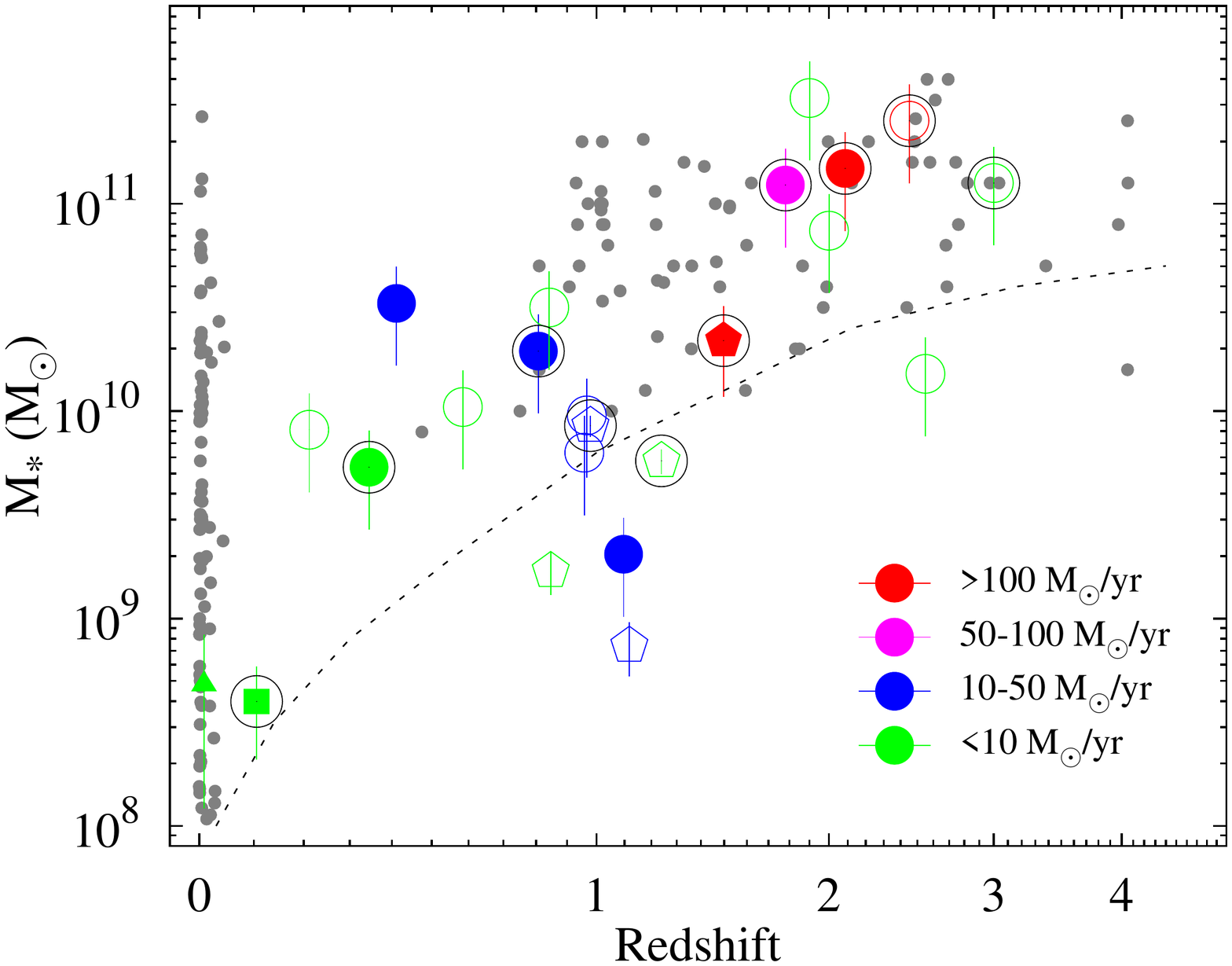}
\caption{Stellar mass against redshift for the sample of GRBHs from this work (pentagons), HPM14 (circles), and for GRBH~980425 \citep[][triangle]{mhp+14} and GRBH~031203 \citep[][square]{sod+14} . The small grey circles correspond to a sample of non-GRBH targets observed and detected with \herschel\ (details in section~\ref{ssec:comp} and Fig.~\ref{fig:dust2star}). Data points are colour-coded according to GRBH SFR. Those GRBHs detected with \herschel\ are plotted as filled symbols, and undetected GRBHs are shown with open symbols. GRBHs that had a GRB with $A_{V,GRB}>1$~mag are indicated with an additional ring drawn around the data point. The dashed line provides a rough divide between those GRBHs detected and undetected with \herschel.}\label{fig:zMstar}
\end{figure}

\section{Discussion}
\label{sec:disc}
The initial aim of our observations had been to study in greater detail the dust properties of GRB host galaxies. In light of our single host galaxy detection, in the following section we look in closer detail at the differences in the galactic properties of those host galaxies that were and were not detected. In order to increase our statistics, in our analysis we include the host galaxy sample from HMP14, and GRBH~980425 and GRBH~031203 from \citet{mhp+14} and \citet{sod+14}. We emphasise that despite the fact that our sample is not complete, all GRBHs included are not `special' within the range of galaxy properties observed in GRBHs (e.g. SGL09, PLT13). Although our compiled sample is on average at the high end of the stellar-mass and SFR distribution (see Fig.~\ref{fig:Hostprops}), there is no compelling reason to believe that these galaxy properties should alter the relation between line of sight dust extinction and galaxy-whole dust emission. For our purposes, our combined sample thus provides a fair representation of dust properties of GRBHs.

\subsection{GRB line of sight versus galaxy-integrated properties}
\label{ssec:AVvsAV}
From the combined sample of GRBHs from this paper, HPM14, \citet{mhp+14} and \citet{sod+14}, a third were detected, and this fraction almost doubles when we only consider those hosts of GRBs with known $A_{V,GRB}>1$~mag\footnotemark[5]. On the other hand, less than 40\% of galaxies hosting so-called dark GRBs (i.e. $\beta_{OX}<0.5$) were detected with \herschel. Although those GRBs in our sample with $A_{V,GRB}>1$~mag are all classified as dark by the $\beta_{OX}$ convention, the converse does not apply, and only 50\% of GRBs classified as dark have $A_V{,GRB}>1$~mag.
This is because although $\beta_{OX}<0.5$ is suggestive of dust extinction, it does not necessarily imply significant amounts of dust, as is the case for $A_{V,GRB}>1$~mag. 
\footnotetext[5]{The visual extinction along the line of sight to GRB~031203 is uncertain due to the large reddening within the Milky Way along the GRB line of sight. However, \citet{pbc+04} estimate $A_{V,GRB}\sim 1$~mag. However, when we apply a more stringent upper limit on $\beta_{OX}$ of $<0.4$ we then select all those GRBs with $A_{V,GRB}>1$~mag, as well as GRBH~970828, for which no $A_{V,GRB}$ information is available. When we only consider the hosts of those GRBs with a measured visual extinction $A_{V,GRB}>1.5$~mag (equivalent to $\beta_{OX}<0.3$) then the detection rate goes up further to three quarters of the sample.}

The fairly high detection rate of hosts selected by their dust-extinguished GRBs implies that the extinguishing dust lies predominantly within the host galaxy ISM, rather than within discrete, dense clouds. If the afterglow extinction arose predominantly from dust within a molecular cloud, either related to or independent of the GRB natal region, then we would expect to have detected a similar number of galaxies with low and high $A_{V,GRB}$. Given the limited number of GRBs with measured $A_{V,GRB}$ we use an optical-to-X-ray spectral index of $\beta_{OX}<0.4$ to identify those GRBs likely to have been significantly dust-extinguished. We find that 50\% of the hosts of dust-extinguished GRBs were detected, whereas the host of only one out of seven relatively unextinguished GRBs was detected with \herschel. Similarly, our results imply that the distribution in $A_{V,GRB}$ is not the result of variations in the host galaxy inclination angle. If by-and-large the host galaxies of GRBs with $A_{V,GRB}>1$~mag were viewed edge-on, and the hosts of relatively unextinguished GRBs were viewed face-on, then we would again expect the \herschel\ detection rate to be independent of $A_{V,GRB}$.

Stellar mass is related to the galaxy dust mass, and thus is clearly an important parameter when considering the galaxy FIR emission. This is illustrated in Fig.~\ref{fig:zMstar}, where we plot the stellar mass as a function of redshift for the combined sample of GRBHs from this work (pentagons) and HPM14 (circle), as well as GRBH~980425 \citep{hpm+14} and GRBH~030325 \citep{sod+14}. For comparison, we also show a sample of non-GRBH galaxies in grey. The apparent trend of increasing stellar mass with higher redshift is a result of selection effects, whereby high-z, low-$M_{\ast}$ galaxies are not generally detected at longer wavelengths. The dashed curve indicates the rough division within the $M_\ast-z$ parameter space between galaxies detected with \herschel\ (including GRBHs), and those undetected GRBHs. Of those GRBHs below this line, 1/5 were detected, whereas above the line, 8/18 were detected. When only considering the hosts of GRBs with $A_{V,GRB}>1$~mag, then only one lies below the curve, and this GRBH was undetected. Above the dashed curve, 6 GRBHs of significantly extinguished GRBs were detected out of 9.

In this figure we also consider the SFR, which is known to correlate with the dust-to-stellar mass ratio \citep[][and references therein]{cee+10, cal01}. Detected GRBHs have progressively higher stellar mass and SFR as redshift increases, which is the combined result of the Malmquist bias and the downsizing of the star formation activity in progressively lower mass galaxies (i.e. the galaxy main sequence). For host galaxies with no FIR detections (open symbols), we use the dust corrected, optically derived SFR (measured from either emission lines or from optical/NIR SED fits), and thus these should be considered as lower limits.

\begin{figure}
\centering
\includegraphics[width=0.57\textwidth]{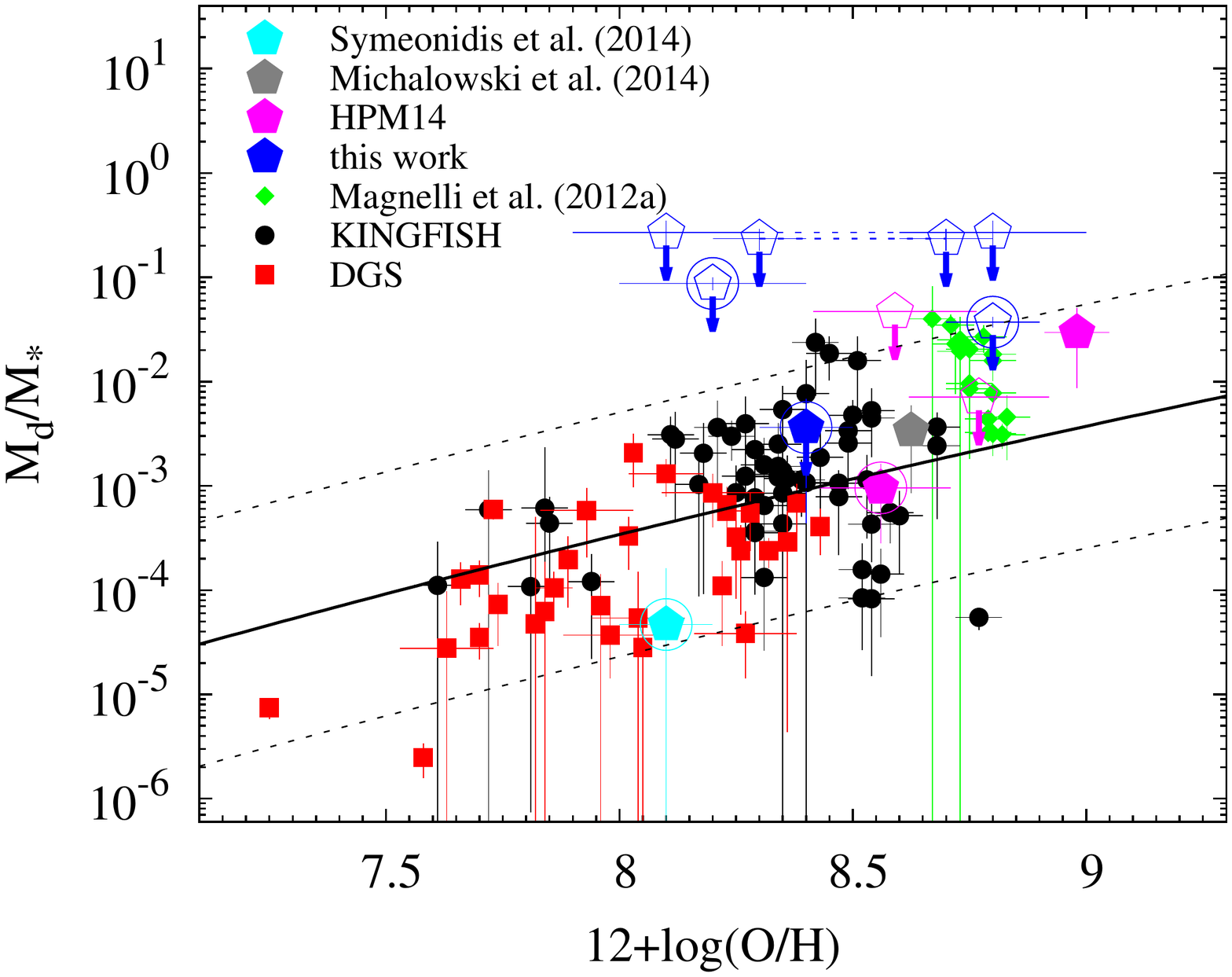}
\caption{$M_d/M_\ast$ as a function of metallicity for a sample of nearby galaxies taken from the KINGFISH \citep[black filled circles;][]{kca+11}, the DGS \citep[red filled squares;][]{mrg+13} \herschel\ guaranteed time key projects, and a sample of high redshift ($z>1$), star forming galaxies from \citet{msl+12} (green filled diamonds). Metallicities of the $z>1$ star forming galaxies were estimated using the mass-metallicity relation and converted into the \citet{dtt02} system. GRBH~031203 \citep[][cyan pentagon]{sod+14}, GRBH~980425 \citep[][grey pentagon]{mhp+14}, and the subset of GRBHs from HPM14 (pink pentagons; GRBHs~980703, 020819B, 050223, 051022) and this paper (blue pentagons) with known metallicity are also plotted, with filled symbols corresponding to \herschel\ detections, and upper limits shown as downward arrows. Those GRBs with afterglow extinction $A_{V,GRB}>1$~mag are indicated with an additional large open circle around the GRBHs. The solid black line is the best-fit power law to the combined KINGFISH and DGS data taken from \citet{rmg+13}, and the dashed lines represent the 3$\sigma$ dispersion.}\label{fig:dust2starZ}
\end{figure}

\begin{figure}
\centering
\includegraphics[width=0.57\textwidth]{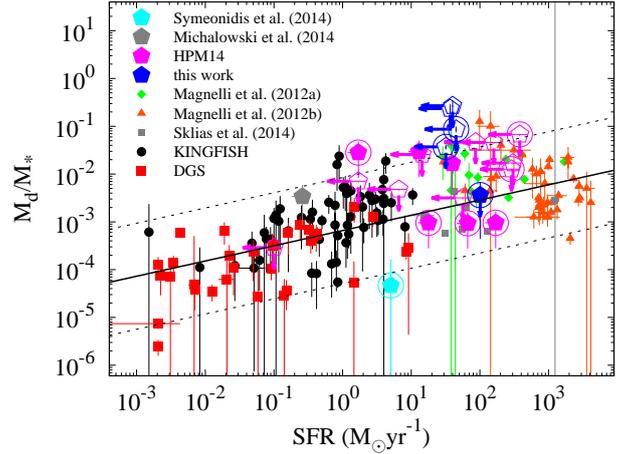}
\caption{$M_d/M_\ast$ as a function of SFR. The same galaxy samples as in Fig.~\ref{fig:dust2starZ} are plotted with the same symbols. Also shown are the sample of star forming lensed galaxies from \citet{szs+14} (small grey squares), and the sample of SMGs from \citet{mls+12} (orange triangles). The solid black line is our best-fit power law to all galaxy samples combined, not including the GRBH sample, and the dashed lines correspond the 3$\sigma$ dispersion.}\label{fig:dust2star}
\end{figure}

Although on the whole using the afterglow line of sight dust extinction is a good diagnostic for identifying GRBs that reside within more massive and dust-rich hosts, there is clear scatter. This is to be expected, given that the light of the afterglow can travel through very different regions of the dusty media within the galaxy-disc plane and/or above it \citep[e.g.][]{ysa+07}. Of those GRBs with $A_{V,GRB}>1$~mag and with host galaxies that were not detected with \herschel, three (GRBHs~071021, 081109, 090926B) have stellar masses that are at the higher end of the GRB host galaxy distribution ($\log[M_\ast/M_\odot]>10$), and they have relatively high SFRs $>10$~M$_\odot$yr$^{-1}$. The non-detection of these galaxies may imply that the large afterglow extinction ($A_{V,GRB}=1.5, 3.4$ and 1.4~mag, respectively) arose predominantly from a dense and fairly isolated dust-cloud, or is associated with a line of sight that crosses the mid-plane of the disk (i.e. with an observed high inclination), rather than from a relatively dust-rich ISM. To place our sample in a broader context, in the next section we compare our results to the dust properties of other samples of galaxies observed with \herschel, covering a range of galaxy types and redshifts.

\subsection{Dust to stellar mass ratio}
\label{ssec:comp}
For our comparison sample, we select galaxies within the local and higher redshift Universe. Within the local Universe, we use the KINGFISH sample, which consists of 61 nearby galaxies of all types \citep{kca+11}, as well as the Dwarf Galaxy Survey \citep[DGS;][]{mrg+13}, which is made up of a sample of 48 galaxies with sub-solar metallicities ($0.03Z_\odot<Z<0.55Z_\odot$). We also include a number of samples at higher redshift to be more comparable to GRBHs. These are a sample of seven strongly lensed galaxies at redshifts $z=1.5-3.2$ \citep{szs+14} taken predominantly from the \herschel\ Lensing Survey \citep[HLS;][]{err+10}, 17 star forming galaxies at $z>1$ \citep{msl+12} that were observed as part of the PACS Evolutionary Probe \citep[PEP;][]{lpa+11} guaranteed time key programme, and a sample of 61 SMGs \citep{mls+12} observed as part of PEP and the \herschel\ Multi-tiered Extragalactic Survey \citep[HerMES,][]{oba+12}. Some properties of these samples are shown in Figs.~\ref{fig:dust2starZ} and \ref{fig:dust2star}.

\citet{rmg+13} combined the sample of nearby galaxies from DGS and KINGFISH \citep{sed+11} and found a positive correlation between the dust-to-stellar mass ratio and galaxy metallicity. In both samples the metallicity was derived using the $R_{23}$ ratio and applying the empirical calibration of \citet{pt05}. In Fig.~\ref{fig:dust2starZ} we reproduce this plot and include the sample of GRBHs from this work (blue), HPM14 (pink)\footnotemark[6], and \citet{sod+14} (cyan) (filled and open pentagons for GRBH detections and upper limits, respectively). Those GRBHs in HPM14 with reported metallicity measurements are GRBHs~980703, 020819B, 050223 and 051022. A range of methods have been used to estimate the metallicities for the GRBHs, depending on the available data. The metallicities for GRBH~000210 and GRBH~000418 \citep{pir14} and GRBH~070306 \citep{ver14} were all derived using the $R_{23}$ ratio and applying the calibration from \citet{kk04}. In the cases where we were unable to select between the lower and higher branch solutions, both metallicities are plotted in Fig.~\ref{fig:dust2starZ} with a dotted line connecting the data points. The metallicities for GRBH~081109 and 090926B \citep{kru+14} were based on a number of line ratio diagnostics from \citet{nmm06}, as were the metallicities for those GRBHs in HPM14 \citep{msc+11}. The line of best-fit to the DGS and KINGFISH galaxies from \citet{rmg+13} (solid line) and 3$\sigma$ dispersion (dashed lines) are plotted as a reference.
\footnotetext[6]{In the case of GRBH~070306, which is present in both this work and in HPM14, we use the values reported in this paper.}

All GRBHs detected with \herschel\ (from this work and from HPM14) have $M_d/M_\ast$ ratios that are within the 3$\sigma$ dispersion of the DGS and KINGFISH samples, and those GRBHs with only upper limits on their dust mass have dust-to-stellar mass ratios that are consistent with the predicted value for their given metallicity. The scatter about the line-of-best-fit shown in Fig.~\ref{fig:dust2starZ} is large, and with our current \herschel\ limits we cannot rule out that the undetected GRBHs have unusually low dust-to-stellar mass ratios. However, the general consistency between the GRBH data points, and the best-fit relation between $M_d/M_\ast$ and $12+\log(O/H)$ implies that the low detection rate in the GRB host galaxy sample is a result of the generally low stellar mass of the sample relative to that of other submm-detected galaxies at similar redshifts.

In Fig.~\ref{fig:dust2star} we plot the dust-to-stellar mass ratio as a function of SFR, where the SFR is derived using FIR data, and for the non-detected GRBHs is therefore an effective upper limit. The data points are the same as in Fig.~\ref{fig:dust2starZ}, but in this figure we include two further high redshift galaxy samples. These are star forming lensed galaxies at redshifts $1.6<z<3.2$ \citep{szs+14} (small grey squares), and a sample of SMGs at redshifts $0.5<z<5.3$ \citep{mls+12} (orange triangles). This plot shows a general trend of increasing dust-to-stellar mass ratio with SFR across all galaxy types, which has previously been observed \citep[][and references therein]{cee+10,cal01}. We fit a linear correlation between the two parameters to the complete sample of galaxies shown in Fig.~\ref{fig:dust2star}, but without including our sample of GRBHs, and found the best fit to be $M_d/M_\ast=0.001\times SFR^{0.3}$ (solid line).

As before, we find that all our non-detected GRBHs have upper limits that lie either above or on the general relation followed by the other galaxy populations. Those GRBHs that were detected all lie within the region of space occupied by the high-z galaxy samples, all of which have $SFR\gtrsim 10$~M$_\odot$yr$^{-1}$, but they have lower SFRs than the majority of the SMG sample. 

Given the dependence between dust temperature and mass, we re-fit all GRBHs assuming the same dust-temperature measured for GRBH~070306 (T=50K) to see how this affected our dust mass upper limits. We found that the dust mass limits decreased in some cases by a factor of 8 or 9. This nevertheless still not does bring the dust-to-mass ratio of our undetected GRBHs below the 3$\sigma$ dispersions shown in Figs.~\ref{fig:dust2starZ} and \ref{fig:dust2star}, and thus our conclusions are broadly unchanged.

\section{Summary}
\label{sec:sum}
We selected a small sample of five GRBHs with evidence of being rich in dust, and used \herschel\ PACS and SPIRE observations to sample the peak of the dust emission within these galaxies. Despite the sensitivity of \herschel, we only detected one GRBH, which had the largest amount of visual extinction along the GRB line of sight and a relatively high stellar mass. In order to improve our statistics, we combined our sample with the GRBHs from HPM14, \citet{mhp+14}, and \citet{sod+14}, all also observed with \herschel. We found a sizeable increase in the \herschel\ detection rate when only considering those hosts of GRBs considerably dust-extinguished afterglows. This implies that the bulk of the afterglow extinguishing dust resides within the ISM of the host galaxy rather than within discrete, dense clouds.

In addition we found that the dust-to-stellar mass ratios and limits of GRBHs are consistent with other star forming galaxy populations selected by different means. Our results thus indicate that the FIR non-detection rate of 60--80~\% within GRBHs is due to the combination of relatively high redshift and low stellar mass of our galaxies. It is possible that metallicity also plays a role, with lower metallicity galaxies of a given stellar mass having lower dust masses than their higher metallicity counterparts. However, a larger number of GRBH metallicity measurements would be needed to investigate this further.

\herschel\ observations of GRBHs have provided the first irrefutable detections of GRBHs at submm wavelengths, and the most accurate sampling of the thermal dust emission peak of these galaxies to date. The full SED coverage provided by optical through to FIR and submm wavelengths with \herschel\ and on-going observatories such as ALMA and JWST enable the properties of GRBHs to be fully characterised, thus resulting in a more complete understanding of the range in environmental properties present within GRBHs.

\acknowledgements
We thank A. R{\'e}my-Ruyer for sharing their best-fit $M_d$ values to the DFG sample, and Francesco Ritacca and the anonymous referee for constructive comments. PS acknowledges support through the Sofja Kovalevskaja Award from the Alexander von Huboldt Foundation of Germany. SS acknowledges support from the Bundesministerium f{\"u}r Wirtschaft and Technologie through DLR (Deutsches Zentrum f{\"u}r Luft- ind Raumfahrt e.V.). LKH and EP are grateful to support from PRIN-INAF 2012/13. DP acknowledges the kind hospitality at the MPE and financial support from the OCEVU LabEx. Part of the funding for GROND (both hardware as well as personel) was granted from the Leibniz-Prize to Prof.~G.~Hasinger (DFG grant HA 1850/28-1). This study is based on data acquired with the Atacama Pathfinder Experiment (APEX), programme ID M-087.F-0024-2011, and the \herschel\ observatory, programme ID OT2\_ppschady\_2. APEX is a collaboration between the Max-Planck-Insitut f{\" u}r Radioastronomie, the European Southern Observatory, and the Onsala Space Observatory. The \herschel\ spacecraft was designed, built, tested, and launched under a contract to ESA managed by the \herschel/Planck Project team by an industrial consortium under the overall responsibility of the prime contractor Thales Alenia Space (Cannes), and including Astrium (Friedrichshafen) responsible for the payload module and for system testing at spacecraft level, Thales Alenia Space (Turin) responsible for the service module, and Astrium (Toulouse) responsible for the telescope, with in excess of a hundred subcontractors. PACS has been developed by a consortium of institutes led by MPE (Germany) and including UVIE (Austria); KU Leuven, CSL, IMEC (Belgium); CEA, LAM (France); MPIA (Germany); INAF-IFSI/OAA/OAP/OAT, LENS, SISSA (Italy); IAC (Spain). This development has been supported by the funding agencies BMVIT (Austria), ESA-PRODEX (Belgium), CEA/CNES (France), DLR (Germany), ASI/INAF (Italy), and CICYT/MCYT (Spain). SPIRE has been developed by a consortium of institutes led by Cardiff University (UK) and including Univ. Lethbridge (Canada); NAOC (China); CEA, LAM (France); IFSI, Univ. Padua (Italy); IAC (Spain); Stockholm Observatory (Sweden); Imperial College London, RAL, UCL-MSSL, UKATC, Univ. Sussex (UK); and Caltech, JPL, NHSC, Univ. Colorado (USA). This development has been supported by national funding agencies: CSA (Canada); NAOC (China); CEA, CNES, CNRS (France); ASI (Italy); MCINN (Spain); SNSB (Sweden); STFC (UK); and NASA (USA). HIPE is a joint development by the \herschel\ Science Ground Segment Consortium, consisting of ESA, the NASA \herschel\ Science Center, and the HIFI, PACS and SPIRE consortia.

\appendix
\onecolumn
\begin{table}
\caption{Optical ($U$ to $Z$) photometric measurements of pre-\swift\ GRB host galaxies. All magnitudes are AB and have been corrected for Galactic reddening \citep{sf11}}\label{tab:OptPhot}
\begin{center}
\begin{tabular}{@{}lcccccc}
\hline
GRB Host & $U$ & $B$ & $V$ & $R$ & $I$ & $Z$ \\
\hline\hline
000210$^{(a)}$ & $24.13^{+0.13}_{-0.13}$ & $24.21^{+0.13}_{-0.13}$ & $24.17^{+0.08}_{-0.08}$ & $23.62^{+0.10}_{-0.10}$ & $22.89^{+0.12}_{-0.12}$ & $23.35^{+0.28}_{-0.28}$ \\
000418$^{(b)}$ & $24.36^{+0.30}_{-0.30}$ & $23.94^{+0.05}_{-0.05}$ & $23.80^{+0.06}_{-0.06}$ & $23.65^{+0.05}_{-0.05}$ & $23.25^{+0.05}_{-0.05}$ & $23.01^{+0.10}_{-0.10}$ \\
\hline
\end{tabular}
\end{center}
Notes: All magnitudes in the AB system and are corrected for Galactic foreground reddening.\\
$^{(a)}$ \citet{gch+03};
$^{(b)}$ \citet{gkc+03}
\end{table}

\begin{table}
\caption{Optical ($u^\prime$ to $z^\prime$) photometric measurements of \swift\ GRB host galaxies}\label{tab:swOptPhot}
\begin{center}
\begin{tabular}{@{}lccccccccc}
\hline
GRB Host & $u^\prime$ & $U$ & $g^\prime$ & $V$ & $r^\prime$ & $R$ & $i^\prime$ & $I$ & $z^\prime$ \\
\hline\hline
070306$^{(a)}$ & $23.05^{+0.46}_{-0.46}$ & $-$ & $22.81^{+0.09}_{-0.09}$ & $-$ & $23.02^{+0.09}_{-0.09}$ & $22.94^{+0.09}_{-0.09}$ & $22.76^{+0.13}_{-0.13}$ & $22.58^{+0.19}_{-0.19}$ & $22.83^{+0.17}_{-0.17}$ \\
081109$^{(b)}$ & $-$ & $23.15^{+0.14}_{-0.14}$ & $23.01^{+0.07}_{-0.07}$ & $22.80^{+0.06}_{-0.06}$ & $22.70^{+0.07}_{-0.07}$ & $-$ & $21.98^{+0.08}_{-0.08}$ & $21.93^{+0.09}_{-0.09}$ & $21.97^{+0.09}_{-0.09}$ \\
090926B$^{(b)}$ & $-$ & $23.61^{+0.13}_{-0.13}$ & $23.23^{+0.07}_{-0.07}$ & $-$ & $22.90^{+0.06}_{-0.06}$ & $-$ & $22.88^{+0.12}_{-0.12}$ & $-$ & $22.41^{+0.10}_{-0.10}$ \\
\hline
\end{tabular}
\end{center}
Notes: All values are as in Table~A\ref{tab:OptPhot}\\
$^{(a)}$ $u^\prime$ and $I$ magnitudes taken from \citet{jrw+08}. All other reported magnitudes taken from \citet{kgs+11};\\
$^{(b)}$ All magnitudes from \citet{kgs+11}
\end{table}

\begin{table}
\caption{NIR ($Y$ to $K$) photometric measurements of GRB host galaxies}\label{tab:NIRPhot}
\begin{center}
\begin{tabular}{@{}lccccccc}
\hline
GRB Host & $Y$ & HST/F110 & $J$ & HST/F125 & HST/F160 & $H$ & $K$ \\
\hline\hline
000210$^{(a)}$ & $-$ & $-$ & $22.90^{+0.10}_{-0.10}$ & $-$ & $-$ &  $22.90^{+0.23}_{-0.23}$ & $22.80^{+0.14}_{-0.14}$ \\
000418$^{(b)}$ & $-$ & $-$ & $23.21^{+0.10}_{-0.10}$ & $-$ & $-$ & $-$ & $23.06^{+0.30}_{-0.30}$ \\
070306$^{(c)}$ & $$ & $-$ & $21.60^{+0.08}_{-0.08}$ & $21.89^{+0.03}_{-0.03}$ & $21.68^{+0.03}_{-0.03}$ & $21.19^{+0.12}_{-0.12}$ & $21.37^{+0.10}_{-0.10}$ \\
081109$^{(c)}$ & $21.61^{+0.08}_{-0.08}$ & $21.50^{+0.03}_{-0.03}$ & $21.36^{+0.06}_{-0.06}$ & $-$ & $21.28^{+0.03}_{-0.03}$ & $21.49^{+0.4}_{-0.4}$ & $21.04^{+0.08}_{-0.08}$ \\
090926B$^{(c)}$ & $-$ & $-$ & $21.86^{+0.13}_{-0.13}$ & $-$ & $-$ & $21.9^{+0.3}_{-0.3}$ & $21.43^{+0.19}_{-0.19}$ \\
\hline
\end{tabular}
\end{center}
Notes: All values are as in Table~A\ref{tab:OptPhot}\\
$^{(a)}$ \citet{gch+03};
$^{(b)}$ \citet{gkc+03};
$^{(c)}$ HST magnitudes taken from \citet{plt+13}. All other reported magnitudes taken from \citet{kgs+11}
\end{table}

\begin{table}
\caption{\spitzer\ photometric measurements of GRB host galaxies}\label{tab:spitPhot}
\begin{center}
\begin{tabular}{@{}l|cccc}
\hline
& \multicolumn{4}{c}{Flux density ($\mu$Jy)}\\
\hline
GRB Host & 3.6$\mu$m & 4.5$\mu$m & 8.0$\mu$m & 24.0$\mu$m \\
\hline\hline
000210$^{(a)}$ & $-$ & $<6.3$ & $15.0\pm 5.1$ & $<31.5$ \\
000418$^{(b)}$ & $-$ & $4.8^{+1.8}_{-1.8}$ & $-$ & $-$ \\
070306$^{(c)}$ & $10.65^{+0.48}_{-0.48}$ & $12.28^{+0.59}_{-0.59}$ & $-$ & $-$ \\
081109$^{(c)}$ & $18.88^{+1.26}_{-1.26}$ & $15.70^{+1.36}_{-1.36}$ & $-$ & $-$ \\
090926B$^{(d)}$ & $10.5^{+0.5}_{-0.5}$ & $7.4^{+0.4}_{-0.4}$ & $-$ & $-$ \\
\hline
\end{tabular}
\end{center}
Notes: Upper limits are given at 3$\sigma$ confidence\\
$^{(a)}$ \citet{mhc+08};
$^{(b)}$ \citet{cmh+10};
$^{(c)}$ \citet{plt+13};
$^{(d)}$ this work
\end{table}

\begin{table}
\caption{Submillimetre photometric measurements of GRB host galaxies}\label{tab:submmPhot}
\begin{center}
\begin{tabular}{@{}lcc}
\hline
& \multicolumn{2}{c}{Flux density (mJy)}\\
GRB Host & 450$\mu$m & 850$\mu$m \\
\hline\hline
000210$^{(a)}$ & $<92.4$ & $2.97^{+0.88}_{-0.88}$ \\
000418$^{(a)}$ & $<56.76$ & $3.15^{+0.90}_{-0.90}$ \\
070306 & $-$ & $<7.44$ \\
081109 & $<13.2$ & $<14.1$ \\
090926B & $-$ & $<14.64$ \\
\hline
\end{tabular}
\end{center}
Notes: Upper limits are given at 3$\sigma$ confidence\\
$^{(a)}$ \citet{bck+03}
\end{table}


\begin{thebibliography}{}
\bibitem[Arnouts et al.(1999)]{acm+99} Arnouts, S., Cristiani, S., Moscardini, L., et al.\ 1999, MNRAS, 310, 540 
\bibitem[Baldry \& Glazebrook(2003)]{bg03} Baldry, I.~K., \& Glazebrook, K.\ 2003, ApJ, 593, 258 
\bibitem[Balog et al.(2013)]{bmn+13} Balog, Z., M{\"u}ller, T., Nielbock, M., et al.\ 2013, Experimental Astronomy, 38 
\bibitem[Berger et al.(2001)]{bdf+01} Berger, E., Diercks, A., Frail, D.~A., et al.\ 2001, ApJ, 556, 556 
\bibitem[Berger et al.(2003)]{bck+03} Berger, E., Cowie, L.~L., Kulkarni, S.~R., et al.\ 2003, ApJ, 588, 99 
\bibitem[Bianchi(2013)]{bia13} Bianchi, S.\ 2013, A\&A, 552, A89 
\bibitem[Bloom et al.(2003)]{bbk+03} Bloom, J.~S., Berger, E., Kulkarni, S.~R., Djorgovski, S.~G., \& Frail, D.~A.\ 2003, AJ, 125, 999 
\bibitem[Bloom et al.(2006)]{bsb+06} Bloom, J.~S., Starr, D.~L., Blake, C.~H., Skrutskie, M.~F., \& Falco, E.~E.\ 2006, Astronomical Data Analysis Software and Systems XV, 351, 751 
\bibitem[Bruzual \& Charlot(2003)]{bc03} Bruzual, G., \& Charlot, S.\ 2003, MNRAS, 344, 1000 
\bibitem[Burrows et al.(2005)]{bhn+05} Burrows, D.~N., Hill, J.~E., Nousek, J.~A., et al.\ 2005, Space Sci. Rev., 120, 165 
\bibitem[Butler et al.(2010)]{bbp10} Butler, N.~R., Bloom, J.~S., \& Poznanski, D.\ 2010, ApJ, 711, 495 
\bibitem[Butler et al.(2012)]{bkf+12} Butler, N., Klein, C., Fox, O., et al.\ 2012, Proc. SPIE, 8446,  
\bibitem[Calzetti et al.(2000)]{cab+00} Calzetti, D., Armus, L., Bohlin, R.~C., et al.\ 2000, ApJ, 533, 682 
\bibitem[Calzetti(2001)]{cal01} Calzetti, D.\ 2001, PASP, 113, 1449 
\bibitem[Casey et al.(2011)]{ccs+11} Casey, C.~M., Chapman, S.~C., Smail, I., et al.\ 2011,MNRAS, 411, 2739 
\bibitem[Casey(2012)]{cas12} Casey, C.~M.\ 2012, MNRAS, 425, 3094 
\bibitem[Castro Cer{\'o}n et al.(2010)]{cmh+10} Castro Cer{\'o}n, J.~M., Micha{\l}owski, M.~J., Hjorth, J., et al.\ 2010, ApJ, 721, 1919 
\bibitem[Chabrier(2003)]{cha03} Chabrier, G.\ 2003, PASP, 115, 763 
\bibitem[da Cunha et al.(2010)]{cee+10} da Cunha, E., Eminian, C., Charlot, S., \& Blaizot, J.\ 2010, MNRAS, 403, 1894 
\bibitem[Dale et al.(2012)]{dae+12} Dale, D.~A., Aniano, G., Engelbracht, C.~W., et al.\ 2012, ApJ, 745, 95 
\bibitem[D'Avanzo et al.(2008)]{dca+08} D'Avanzo, P., Covino, S., Antonelli, L.~A., et al.\ 2008, GRB Coordinates Network, 8501, 1 
\bibitem[Denicol{\'o} et al.(2002)]{dtt02} Denicol{\'o}, G., Terlevich, R., \& Terlevich, E.\ 2002, MNRAS, 330, 69 
\bibitem[Dunne \& Eales(2001)]{de01} Dunne, L., \& Eales, S.~A.\ 2001, MNRAS, 327, 697 
\bibitem[Egami et al.(2010)]{err+10} Egami, E., Rex, M., Rawle, T.~D., et al.\ 2010, A\&A, 518, L12 
\bibitem[Elliott et al.(2013)]{ekg+13} Elliott, J., Kr{\"u}hler, T., Greiner, J., et al.\ 2013, A\&A, 556, A23 
\bibitem[Fruchter et al.(2006)]{fls+06} Fruchter, A.~S., Levan, A.~J., Strolger, L., et al.\ 2006, Nature, 441, 463 
\bibitem[Fynbo et al.(2009)]{fmj+09} Fynbo, J.~P.~U., Malesani, D., Jakobsson, P., \& D'Elia, V.\ 2009, GRB Coordinates Network, 9947, 1 
\bibitem[Galama et al.(1999)]{gvp+99} Galama, T.~J., Vreeswijk, P.~M., van Paradijs, J., et al.\ 1999, A\&A, 138, 465 
\bibitem[Gehrels et al.(2004)]{gcg+04} Gehrels, N., Chincarini, G., Giommi, P., et al.\ 2004, \apj, 611, 1005 
\bibitem[Gorosabel et al.(2003a)]{gch+03} Gorosabel, J., Christensen, L., Hjorth, J., et al.\ 2003a, A\&A, 400, 127 
\bibitem[Gorosabel et al.(2003b)]{gkc+03} Gorosabel, J., Klose, S., Christensen, L., et al.\ 2003b, A\&A, 409, 123 
\bibitem[Graham et al.(2009)]{gfk+09} Graham, J.~F., Fruchter, A.~S., Kewley, L.~J., et al.\ 2009, American Institute of Physics Conference Series, 1133, 269
\bibitem[Graham \& Fruchter(2013)]{gf13} Graham, J.~F., \& Fruchter, A.~S.\ 2013, ApJ, 774, 119 
\bibitem[Greiner et al.(2008)]{gbc+08} Greiner, J., Bornemann, W., Clemens, C., et al.\ 2008, PASP, 120, 405 
\bibitem[Greiner et al.(2011)]{gkk+11} Greiner, J., Kr{\"u}hler, T., Klose, S., et al.\ 2011, A\&A, 526, A30 
\bibitem[Griffin et al.(2010)]{gaa+10} Griffin, M.~J., Abergel, A., Abreu, A., et al.\ 2010, A\&A, 518, L3 
\bibitem[G{\"u}sten et al.(2006)]{gns+06} G{\"u}sten, R., Nyman, L.~{\AA}., Schilke, P., et al.\ 2006, A\&A, 454, L13 
\bibitem[Hjorth et al.(2003)]{hsm+03} Hjorth, J., Sollerman, J., M{\o}ller, P., et al.\ 2003, Nature, 423, 847 
\bibitem[Hjorth et al.(2012)]{hmj+12} Hjorth, J., Malesani, D., Jakobsson, P., et al.\ 2012, ApJ, 756, 187 
\bibitem[Hjorth et al.(2014)]{hgm+14} Hjorth, J., Gall, C., \& Micha{\l}owski, M.~J.\ 2014, ApJ, 782, L23 
\bibitem[Holland et al.(1999)]{hrg+99} Holland, W.~S., Robson, E.~I., Gear, W.~K., et al.\ 1999, MNRAS, 303, 659 
\bibitem[Hunt et al.(2011)]{hpr+11} Hunt, L., Palazzi, E., Rossi, A., et al.\ 2011, ApJ, 736, L36 
\bibitem[Hunt et al.(2014)]{hpm+14} Hunt, L., Palazzi, E., Micha{\l}owski, M.~J., et al.\ 2014, A\&A, in press (HPM14) 
\bibitem[Ilbert et al.(2006)]{iam+06} Ilbert, O., Arnouts, S., McCracken, H.~J., et al.\ 2006, A\&A, 457, 841 
\bibitem[Ilbert et al.(2010)]{isl+10} Ilbert, O., Salvato, M., Le Floc'h, E., et al.\ 2010, ApJ, 709, 644 
\bibitem[Jakobsson et al.(2004)]{jhf+04} Jakobsson, P., Hjorth, J., Fynbo, J.~P.~U., et al.\ 2004, ApJ, 617, L21 
\bibitem[Jaunsen et al.(2008)]{jrw+08} Jaunsen, A.~O., Rol, E., Watson, D.~J., et al.\ 2008, ApJ, 681, 453 
\bibitem[Kajisawa et al.(2009)]{kit+09} Kajisawa, M., Ichikawa, T., Tanaka, I., et al.\ 2009, ApJ, 702, 1393 
\bibitem[Kelly et al.(2013)]{kff+13} Kelly, P.~L., Filippenko, A.~V., Fox, O.~D., Zheng, W., \& Clubb, K.~I.\ 2013, ApJ, 775, L5 
\bibitem[Kelly et al.(2014)]{kfm+14} Kelly, P.~L., Filippenko, A.~V., Modjaz, M., \& Kocevski, D.\ 2014, submitted to ApJ (arXiv:1401.0729)
\bibitem[Kennicutt(1998)]{ken98} Kennicutt, R.~C., Jr.\ 1998, Ann. Rev. A\&A, 36, 189 
\bibitem[Kennicutt et al.(2011)]{kca+11} Kennicutt, R.~C., Calzetti, D., Aniano, G., et al.\ 2011, PASP, 123, 1347 
\bibitem[Kewley \& Ellison(2008)]{ke08} Kewley, L.~J., \& Ellison, S.~L.\ 2008, ApJ, 681, 1183 
\bibitem[Kistler et al.(2009)]{kyb+09} Kistler, M.~D., Y{\"u}ksel, H., Beacom, J.~F., Hopkins, A.~M., \& Wyithe, J.~S.~B.\ 2009, ApJ, 705, L104 
\bibitem[Klose et al.(2000)]{ksm+00} Klose, S., Stecklum, B., Masetti, N., et al.\ 2000, ApJ, 545, 271 
\bibitem[Kobulnicky \& Kewley(2004)]{kk04} Kobulnicky, H.~A., \& Kewley, L.~J.\ 2004, ApJ, 617, 240 
\bibitem[Kocevski \& West(2011)]{kw11} Kocevski, D., \& West, A.~A.\ 2011, ApJ, 735, L8 
\bibitem[Kr{\"u}hler et al.(2011)]{kgs+11} Kr{\"u}hler, T., Greiner, J., Schady, P., et al.\ 2011, A\&A, 534, A108 
\bibitem[Kr{\"u}hler et al.(2014)]{kru+14} Kr{\"u}hler, T., et al.\ 2014, in preparation  
\bibitem[K{\"u}pc{\"u} Yolda{\c s} et al.(2007)]{ysa+07} K{\"u}pc{\"u} Yolda{\c s}, A., Salvato, M., Greiner, J., et al.\ 2007, A\&A, 463, 893 
\bibitem[Langer \& Norman(2006)]{ln06} Langer, N., \& Norman, C.~A.\ 2006, ApJ, 638, L63 
\bibitem[Levesque et al.(2010b)]{lkb+10} Levesque, E.~M., Kewley, L.~J., Berger, E., \& Zahid, H.~J.\ 2010a, AJ, 140, 1557 
\bibitem[Levesque et al.(2010b)]{lkg+10} Levesque, E.~M., Kewley, L.~J., Graham, J.~F., \& Fruchter, A.~S.\ 2010b, ApJ, 712, L26 
\bibitem[Lo Faro et al.(2013)]{lfv+13} Lo Faro, B., Franceschini, A., Vaccari, M., et al.\ 2013, ApJ, 762, 108 
\bibitem[Lutz et al.(2011)]{lpa+11} Lutz, D., Poglitsch, A., Altieri, B., et al.\ 2011, A\&A, 532, A90 
\bibitem[MacFadyen \& Woosley(1999)]{mw99} MacFadyen, A.~I., \& Woosley, S.~E.\ 1999, ApJ, 524, 262 
\bibitem[Madden et al.(2013)]{mrg+13} Madden, S.~C., R{\'e}my-Ruyer, A., Galametz, M., et al.\ 2013, PASP, 125, 600 
\bibitem[Magdis et al.(2012)]{mdb+12} Magdis, G.~E., Daddi, E., B{\'e}thermin, M., et al.\ 2012, ApJ, 760, 6 
\bibitem[Magdis et al.(2013)]{mrh+13} Magdis, G.~E., Rigopoulou, D., Helou, G., et al.\ 2013, A\&A, 558, A136 
\bibitem[Magnelli et al.(2012a)]{msl+12} Magnelli, B., Saintonge, A., Lutz, D., et al.\ 2012a, A\&A, 548, A22 
\bibitem[Magnelli et al.(2012b)]{mls+12} Magnelli, B., Lutz, D., Santini, P., et al.\ 2012b, A\&A, 539, A155 
\bibitem[Malesani et al.(2009)]{mfd09} Malesani, D., Fynbo, J.~P.~U., \& D'Elia, V.\ 2009, GRB Coordinates Network, 9944, 1 
\bibitem[Mannucci et al.(2011)]{msc+11} Mannucci, F., Salvaterra, R., \& Campisi, M.~A.\ 2011, MNRAS, 414, 1263 
\bibitem[Micha{\l}owski et al.(2008)]{mhc+08} Micha{\l}owski, M.~J., Hjorth, J., Castro Cer{\'o}n, J.~M., \& Watson, D.\ 2008, ApJ, 672, 817 
\bibitem[Micha{\l}owski et al.(2010)]{mhw10} Micha{\l}owski, M., Hjorth, J., \& Watson, D.\ 2010, A\&A, 514, A67 
\bibitem[Micha{\l}owski et al.(2012a)]{mdc+12} Micha{\l}owski, M.~J., Dunlop, J.~S., Cirasuolo, M., et al.\ 2012a, A\&A, 541, A85 
\bibitem[Micha{\l}owski et al.(2012b)]{mkh+12} Micha{\l}owski, M.~J., Kamble, A., Hjorth, J., et al.\ 2012b, ApJ, 755, 85 
\bibitem[Micha{\l}owski et al.(2014)]{mhp+14} Micha{\l}owski, M.~J., Hunt, L.~K., Palazzi, E., et al.\ 2014, A\&A, 562, A70 
\bibitem[Mitchell et al.(2013)]{mlb+13} Mitchell, P.~D., Lacey, C.~G., Baugh, C.~M., \& Cole, S.\ 2013, MNRAS, 435, 87 
\bibitem[Modjaz et al.(2008)]{mkk+08} Modjaz, M., Kewley, L., Kirshner, R.~P., et al.\ 2008, AJ, 135, 1136 
\bibitem[Nagao et al.(2006)]{nmm06} Nagao, T., Maiolino, R., \& Marconi, A.\ 2006, A\&A, 459, 85 
\bibitem[Oliver et al.(2012)]{oba+12} Oliver, S.~J., Bock, J., Altieri, B., et al.\ 2012, MNRAS, 424, 1614 
\bibitem[Ott(2010)]{ott10} Ott, S.\ 2010, Astronomical Data Analysis Software and Systems XIX, 434, 139 
\bibitem[Pearson et al.(2013)]{pln+13} Pearson, C., Lim, T., North, C., et al.\ 2013, Experimental Astronomy, 37 
\bibitem[Peng et al.(2002)]{phi+02} Peng, C.~Y., Ho, L.~C., Impey, C.~D., \& Rix, H.-W.\ 2002, AJ, 124, 266 
\bibitem[P{\'e}rez-Gonz{\'a}lez et al.(2010)]{per+10} P{\'e}rez-Gonz{\'a}lez, P.~G., Egami, E., Rex, M., et al.\ 2010, A\&A, 518, L15 
\bibitem[Perley et al.(2013)]{plt+13} Perley, D.~A., Levan, A.~J., Tanvir, N.~R., et al.\ 2013, ApJ, 778, 128 (PLT13)
\bibitem[Perley \& Perley(2013)]{pp13} Perley, D.~A., Perley, R.~A.\ 2013, ApJ, 778, 172 
\bibitem[Pilbratt et al.(2010)]{prp+10} Pilbratt, G.~L., Riedinger, J.~R., Passvogel, T., et al.\ 2010, A\&A, 518, L1 
\bibitem[Pilyugin \& Thuan(2005)]{pt05} Pilyugin, L.~S., \& Thuan, T.~X.\ 2005, ApJ, 631, 231 
\bibitem[Piranomonte et al.(2014)]{pir14} Piranomonte, S., et al.\ 2014, in progress 
\bibitem[Piro et al.(2002)]{pfg+02} Piro, L., Frail, D.~A., Gorosabel, J., et al.\ 2002, ApJ, 577, 680 
\bibitem[Poglitsch et al.(2010)]{pwg+10} Poglitsch, A., Waelkens, C., Geis, N., et al.\ 2010, A\&A, 518, L2 
\bibitem[Pozzetti et al.(2007)]{pbl+07} Pozzetti, L., Bolzonella, M., Lamareille, F., et al.\ 2007, A\&A, 474, 443 
\bibitem[Prochaska et al.(2004)]{pbc+04} Prochaska, J.~X., Bloom, J.~S., Chen, H.-W., et al.\ 2004, ApJ, 611, 200 
\bibitem[R{\'e}my-Ruyer et al.(2013)]{rmg+13} R{\'e}my-Ruyer, A., Madden, S.~C., Galliano, F., et al.\ 2013, A\&A, 557, A95 
\bibitem[Robertson \& Ellis(2012)]{re12} Robertson, B.~E., \& Ellis, R.~S.\ 2012, ApJ, 744, 95 
\bibitem[Rol et al.(2007)]{rlt+07} Rol, E., Levan, A., Tanvir, N., Schirmer, M., \& Castro-Tirado, A.~J.\ 2007, GRB Coordinates Network, 6174, 1 
\bibitem[Rossi et al.(2012)]{rks+12} Rossi, A., Klose, S., Ferrero, P., et al.\ 2012, A\&A, 545, A77 
\bibitem[Salpeter(1955)]{sal55} Salpeter, E.~E.\ 1955, ApJ, 121, 161 
\bibitem[Savaglio et al.(2005)]{sgl+05} Savaglio, S., Glazebrook, K., Le Borgne, D., et al.\ 2005, ApJ, 635, 260 
\bibitem[Savaglio et al.(2009)]{sgl09} Savaglio, S., Glazebrook, K., \& Le Borgne, D.\ 2009, ApJ, 691, 182 (SGL09)
\bibitem[Savaglio et al.(2012)]{srg+12} Savaglio, S., Rau, A., Greiner, J., et al.\ 2012, MNRAS, 420, 627 
\bibitem[Schady et al.(2007)]{smp+07} Schady, P., Mason, K.~O., Page, M.~J., et al.\ 2007, MNRAS, 377, 273 
\bibitem[Schady et al.(2012)]{sdp+12} Schady, P., Dwelly, T., Page, M.~J., et al.\ 2012, A\&A, 537, A15 
\bibitem[Schlafly \& Finkbeiner(2011)]{sf11} Schlafly, E.~F., \& Finkbeiner, D.~P.\ 2011, ApJ, 737, 103 
\bibitem[Silva et al.(1998)]{sgb+98} Silva, L., Granato, G.~L., Bressan, A., \& Danese, L.\ 1998, ApJ, 509, 103
\bibitem[Siringo et al.(2009)]{skk+09} Siringo, G., Kreysa, E., Kov{\'a}cs, A., et al.\ 2009, A\&A, 497, 945 
\bibitem[Skibba et al.(2011)]{sed+11} Skibba, R.~A., Engelbracht, C.~W., Dale, D., et al.\ 2011, ApJ, 738, 89 
\bibitem[Sklias et al.(2014)]{szs+14} Sklias, P., Zamojski, M., Schaerer, D., et al.\ 2014, A\&A, 561, A149 
\bibitem[Svensson et al.(2010)]{slt+10} Svensson, K.~M., Levan, A.~J., Tanvir, N.~R., Fruchter, A.~S., \& Strolger, L.-G.\ 2010, MNRAS, 405, 57 
\bibitem[Svensson et al.(2012)]{slt+12} Svensson, K.~M., Levan, A.~J., Tanvir, N.~R., et al.\ 2012, MNRAS, 421, 25 
\bibitem[Symeonidis et al.(2014)]{sod+14} Symeonidis, M., Oates, S.~R., de Pasquale, M., et al.\ 2014, MNRAS, in press (arXiv:1406.2599) 
\bibitem[Tanvir et al.(2004)]{ttb+04} Tanvir, N.~R., Barnard, V.~E., Blain, A.~W., et al.\ 2004, MNRAS, 352, 1073 
\bibitem[de Ugarte Postigo et al.(2012)]{ulm+12} de Ugarte Postigo, A., Lundgren, A., Mart{\'{\i}}n, S., et al.\ 2012, A\&A, 538, A44 
\bibitem[Vergani et al.(2014)]{ver14} Vergani, S.~D., et al.\ 2014, in progress 
\bibitem[Wanderman \& Piran(2010)]{wp10} Wanderman, D., \& Piran, T.\ 2010, MNRAS, 406, 1944 
\bibitem[Wolf \& Podsiadlowski(2007)]{wp07} Wolf, C., \& Podsiadlowski, P.\ 2007, MNRAS, 375, 1049 
\bibitem[Wuyts et al.(2009)]{wfc+09} Wuyts, S., Franx, M., Cox, T.~J., et al.\ 2009, ApJ, 696, 348 
\end{thebibliography}
\end{document}